%% file: main.tex
\documentclass[preprint]{elsarticle} 

\makeatletter
	\def\ps@pprintTitle{%
 	\let\@oddhead\@empty
	\let\@evenhead\@empty
	\def\@oddfoot{\centerline{\thepage}}%
	\let\@evenfoot\@oddfoot}
\makeatother
\biboptions{square,numbers,sort&compress}
\usepackage[textsize=tiny]{todonotes}
\usepackage{etoolbox}
\patchcmd{\MaketitleBox}{\footnotesize\itshape\elsaddress\par\vskip36pt}{\footnotesize\itshape\elsaddress\par\parbox[b][36pt]{\linewidth}{\vfill\hfill\textnormal{\today}\hfill\null\vfill}}{}{}%
\patchcmd{\pprintMaketitle}{\footnotesize\itshape\elsaddress\par\vskip36pt}{\footnotesize\itshape\elsaddress\par\parbox[b][36pt]{\linewidth}{\vfill\hfill\textnormal{\today}\hfill\null\vfill}}{}{}%

\usepackage[colorlinks=true,%
            breaklinks=true,%
            linkcolor=blue,
            urlcolor=blue,%
            citecolor=blue,%
            pdftitle={Title of paper}, 
            pdfkeywords={Keywords},	
            pdfauthor={Authors},		
            bookmarksopen=false,
            pdfpagemode=UseNone]{hyperref}

\input{ListPackages}
\usepackage{siunitx}
\makeatletter
\renewcommand*\env@matrix[1][\arraystretch]{%
  \edef\arraystretch{#1}%
  \hskip -\arraycolsep
  \let\@ifnextchar\new@ifnextchar
  \array{*\c@MaxMatrixCols c}}
\makeatother

\input{headerCommands.tex}
\input{headerCommandsAlma.tex}

\begin{document}
\begin{frontmatter}
    \title{Parallel diffusion operator for magnetized plasmas with improved spectral fidelity}
    \author[ga]{Federico D. Halpern\corref{cor1}}\ead{halpernf@fusion.gat.com}
    \cortext[cor1]{Corresponding author}
    \author[ga]{Min-Gu Yoo}
    \author[ga]{Brendan Lyons}
    \author[ga]{Juan Diego Colmenares}
    \address[ga]{General Atomics, P.O. Box 85608, San Diego, CA, USA}
    
    \begin{abstract}
    Diffusive transport processes in magnetized plasmas are highly anisotropic, with fast parallel transport along the magnetic field lines sometimes faster than perpendicular transport by orders of magnitude. This constitutes a major challenge for describing non-grid-aligned magnetic structures in Eulerian (grid-based) simulations. The present paper describes and validates a new method for parallel diffusion in magnetized plasmas based on the anti-symmetry representation [Halpern and Waltz, Phys.~Plasmas 25, 060703 (2018)]. In the anti-symmetry formalism, diffusion manifests as a flow operator involving the logarithmic derivative of the transported quantity. Qualitative plane wave analysis shows that the new operator naturally yields better discrete spectral resolution compared to its conventional counterpart. Numerical simulations comparing the new method against existing finite difference methods are carried out, showing significant improvement. In particular, we find that combining anti-symmetry with finite differences in diagonally staggered grids essentially eliminates the so-called ``artificial numerical diffusion'' that affects conventional finite difference and finite volume methods.
    \end{abstract}	
    \begin{keyword}
    Anisotropic diffusion \sep finite difference methods \sep conservation laws \sep anti-symmetry 
    \end{keyword}
\end{frontmatter}


\section{Introduction}\label{sec:Intro}
Magnetized plasmas are characterized by dynamics that are strongly anisotropic with respect to the local magnetic field line pitch angle. In the collisional regime \cite{Braginskii1965}, in particular, the heat conductivity parallel to the magnetic field acts in a very short time scale, with heat conductivity ratios $\kappa_\parallel/\kappa_\perp$ sometimes $10^9$ or larger. Hence, gradients parallel to the magnetic field are much shallower than gradients perpendicular to it. For a typical field aligned perturbation, the parallel scale length is macroscopic -- of the order of the magnetic field line length -- while the typical perpendicular length scale is microscopic -- of the order of the ion sound Larmor gyroradius $\rho_s=\sqrt{mT}/(eB)$. This paper concentrates on numerical strategies for calculating the fast parallel heat diffusion with high accuracy and fidelity in numerical simulations.

There are several interrelated motives that make it necessary to properly satisfy the field-aligned property. First and foremost, satisfying the this property is crucial to obtain the correct dynamics in simulations. Second, field-aligned dynamics allow a reduction of the simulation grid resolution in the direction parallel (or nearly parallel) to the magnetic field, drastically reducing simulation cost. Third, since the parallel dynamics act in a short timescale, reducing the resolution also relaxes time stepping constraints.

Eulerian (grid-based) simulations must typically overcome one key obstacle to properly describe parallel heat transport. Small discretization errors produced when projecting the dynamics onto the parallel direction manifest, instead, in the perpendicular direction. If the parallel diffusion is erroneously projected along the perpendicular axis, it causes artificial, numerical diffusion of the perpendicular gradients. Since the parallel diffusion coefficients are large, a small projection error can cause significant unphysical damping of the perpendicular dynamics. The phenomenon is informally called ``artificial perpendicular diffusion,'' and it still constitutes a major challenge in plasma simulations. 

Plasma codes have adopted different strategies to exploit the anisotropy between parallel and perpendicular dynamics while minimizing artificial diffusion. Two principal families of strategies have emerged. In the first strategy, the plasma equations are expressed and discretized using magnetic field line following coordinates~\cite{Connor1978, Beer1995}. The parallel heat diffusion operator becomes one dimensional in discrete space, with trivial implementation and high accuracy. This strategy is currently employed in most gyrokinetic codes~\cite{Jenko2000, Candy2003, Jolliet2007, Candy2016, Hakim2020} and in some drift-fluid and gyrofluid codes as well~\cite{Scott1997,Dudson2009,Dudson2024}. There are two main drawabacks to this strategy. First, they make self-consistent evolution of the magnetic geometry challenging. Second, field aligned methods are challenging to use at or near magnetic nulls (X-points). The coordinate system becomes ill-defined at the X-point, with the Jacobian of the transformation and some of the metric coefficients becoming poorly behaved as the null is approached. This issue is usually circumvented by utilizing a multi-block approach or by avoiding the (singular) X-point location itself (\emph{e.g.}~\cite{Dudson2024}), which nevertheless remains numerically challenging.

In the second strategy, the plasma quantities are expressed in a non-field-aligned grid, but a high-order discretization method is used. In principle, this allows for arbitrary magnetic configurations to be modeled if the discrete operators preserve enough physics fidelity. This strategy is often employed in extended magnetohydrodynamics (MHD) codes~\cite{Chacon2008, Lutjens2010, Todo2010, Jardin2012}, which evolve the magnetic equilibrium fields in time, and in drift-fluid or gyrofluid codes~\cite{Hariri2013, Tamain2016, Paruta2018, Zhu2018, Stegmeir2018}. In extended MHD codes, the fast parallel diffusion plays a key role describing tearing mode onset (see~\cite{LaHaye2006}), thus high-order finite element or spectral methods are preferred. One drawback in these methods is that they are spatially non-local, and thus not optimally suited for massive paralellization. Precision and fidelity are obtained at the cost of computational expense and simulation wall clock time.

Other authors have developed stencil-based techniques for non-field-aligned grids, which instead prioritize data locality in order to minimize computational cost and maximize parallel scalability. These techniques are more appropriate for global drift-fluid and gyro-fluid turbulence simulations, where the impact of parallel heat fluxes is important, but less sensitive than in extended MHD. Hariri and Ottaviani~\cite{Ottaviani2011, Hariri2013} developed the Flux-Coordinate Independent (FCI) method, which involves a transform of the parallel dynamics operators to locally aligned coordinates. The operators are evaluated using a cubic Hermite reconstruction of the field values where the magnetic field lines cross the numerical grid. The technique allows implementation of arbitrary magnetic field configurations, including magnetic nulls~\cite{Hariri2015}, and is currently in use in both fluid and gyrokinetic codes~\cite{Stegmeir2018, Zhu2018, Wiesenberger2019, Michels2021}. 


It is also possible to evaluate the parallel diffusion using a non-field aligned projection of the Laplacian operator using high-order finite difference or finite volume discretization, \emph{e.g.}~\cite{Tamain2015, Paruta2018, Dorf2021, Giacomin2022, Yoo2024}. This approach is less commonly invoked in the literature, principally because Cartesian or cylindrical decomposition of the Laplacian can introduce significant perpendicular artificial diffusion, potentially increasing the resolution requirements. For instance, a recent paper \cite{Stegmeir2023} compares the FCI method against a classic finite difference parallel diffusion method \cite{Shashkov1996, Guenter2005, Guenter2007}, concluding that the FCI method has beneficial properties in fidelity, required grid sizes, and time-stepping costs.

Nevertheless, a suitable finite difference method still presents several potential advantages. Not only does it allow arbitrary magnetic configurations to be modeled, but it also allows a time-varying equilibrium magnetic field, without incurring in any additional costs. This is unlike the FCI method, which requires a costly re-evaluation of the interpolation basis as the magnetic field line pitch varies. Finite differences also carry, potentially, a small computational cost, algorithmic flexibility (\emph{e.g.} the ability to use flux limiters coupled to high-order finite volume methods~\cite{Liu1994, Dorf2021}), and it is trivially parallelizable and vectorizable. The latter point is a key factor required to develop a powerful exascale extended MHD or global turbulence code addressing reactor scale at high resolution. 

In this manuscript, we make a novel contribution towards arbitrary non-field-aligned discretizations using simple finite difference methods. More specifically, we present a new formulation of the parallel diffusion operator based on the anti-symmetry paradigm~\cite{Halpern2018}. Anti-symmetry operates by carrying out a transformation of the plasma equations to a frame where the dynamics constitute infinitesimal rotations. In this frame, the dynamics result from a completely anti-symmetric force operator acting on a vector state. Using the anti-commutative property $\psi^T\mat{F}\phi=-\phi\mat{F}\psi$ of the force operator $\mat{F}$, exact conservation and numerical stability are obtained by construction even when using simple stencil methods. Physical observables are conceptualized as square quantities, and auxiliary variables are advanced in time. These variables are related to the square root of the observables. This means that profile gradients are mapped to a lower wave number, possibly increasing the effective resolution in numerical simulations.

So far, the discussions on anti-symmetry method have concentrated on the reformulation and verification of the advection (hyperbolic) component of the standard fluid equations~\cite{Halpern2020, Halpern2021}. This was carried out under the assumption that a suitable, Onsager-symmetric collision operator would be provided. This manuscript presents the first advance towards better describing diffusive terms in a manner consistent with the rest of the approach.

Our main results are as follows. First, we present a simple derivation of the anti-symmetry parallel diffusion operator, and discuss two potential discretization strategies. One of the discretizations is akin to a finite volume method, while the other one follows a similar recipe as the one employed in \cite{Shashkov1996}. We estimate the discrete operator spectral response function for plane wave dynamics, and we compare it to existing algorithms. This analysis indicates that anti-symmetry based methods can have better spectral accuracy, \emph{e.g.} comparing the spectral response of the diffusion operator to its Fourier analog. In practical terms, this decreases the artificial numerical diffusion incurred in numerical simulations.

As a side effect, and owing to the conceptual simplicity of the methods used, we are able to offer comprehensive analysis and measurements of the artificial perpendicular diffusion effect, which we can compare against traditional stencil-based discretizations of the parallel Laplacian. The numerical properties of the new approach are demonstrated in 2D and 3D simulations of increasing complexity, including a 1-to-1 comparisons with tests proposed in \cite{Stegmeir2023}. While all the algorithms tested (conventional and anti-symmetry) converge following their formal precision order used (2nd, 4th, and 6th), the anti-symmetry method produces smaller error and smaller artificial diffusion than its counterparts. We conclude that the new anti-symmetry method results in significantly reduced artificial perpendicular diffusion compared to traditional approaches.

The rest of the paper is organized as follows. Section~\ref{sec:Algorithms} briefly introduces an anti-symmetry approach for parallel diffusion, and describes its numerical discretization. Section~\ref{sec:Response} discusses the spectral properties of each parallel diffusion discrete scheme. Section~\ref{sec:Simulations} describes numerical simulations of the algorithms provided in the previous section, including an analysis of the artificial perpendicular diffusion introduced by the numerical schemes. Some concluding remarks are provided in Sec.~\ref{sec:Conclusion}.

\section{Anti-symmetry diffusion operator and finite difference discretization strategy}\label{sec:Algorithms}
The purpose of this section is to provide an alternative, anti-symmetry-like representation of the diffusion operator as well as two different discretization strategies based on finite differences. To derive the anti-symmetry diffusion operator, we start from a simple diffusion equation $\partial_t f = \nabla\cdot(\tensor{K}\cdot\nabla f)$, where $\tensor{K}$ is an arbitrary diffusion tensor. Assuming that $f$ is a positive scalar, we notice that the equation can be rewritten in terms of $\sqf$ and its logarithmic gradient:
\begin{align}
    \ddt{\sqf^2}{t} & = \sqf\left(\nabla\cdot\vec{v}_{\mathbf{K}}+\vec{v}_{\mathbf{K}}\cdot\nabla\right)\sqf,\label{eq:AntiDiff}\\
    \vec{v}_{\mathbf{K}} & = 2\tensor{K}\cdot\nabla (\log \sqf).
\end{align}(The shorthand notation $\left(\nabla\cdot\vec{v}_\mathbf{K}+\vec{v}_\mathbf{K}\cdot\nabla\right)\sqf$ implies $\divp{\vec{v}_\mathbf{K}\sqf}+\vec{v}_\mathbf{K}\cdot\nabla\sqf$.) The original diffusion equation has been recast as a flow process, which is driven by the projection of the diffusion tensor onto the logarithmic gradient of the transported quantity. The physical quantity $f=(\sqf)^2$ is conserved, which can be easily demonstrated from the volume integral of Eq.~\ref{eq:AntiDiff} together with the anti-symmetry property (up to surface terms) of the flow operator $\int dV \sqf\left(\nabla\cdot\vec{v}_\mathbf{K}+\vec{v}_\mathbf{K}\cdot\nabla\right)\sqf = -\int dV \sqf\left(\nabla\cdot\vec{v}_\mathbf{K}+\vec{v}_\mathbf{K}\cdot\nabla\right)\sqf$.

In numerical applications, the left-hand-side of Eq.~\ref{eq:AntiDiff} is rewritten in terms of $\sqf$, and $\sqf$ is evolved instead of $f$. In the following, the left-hand-side is retained \emph{as-is} as a visual clue that we are investigating the properties of the right-hand-side on $f$ instead of $\sqf$. A discussion of discretization strategies using finite differences now follows. 

\subsection{Discretization strategy}
In the following, we employ finite differences using staggered grids in standard Cartesian coordinates. This is a departure from the centered finite difference methods presented in \cite{Halpern2021}. Two different strategies are devised, but more may exist. Both strategies rely on defining consistent gradient and divergence operators such that their successive application is equivalent to applying the Laplace operator. Extension to general coordinates is possible using curvilinear transform of Eq.~\ref{eq:AntiDiff} \cite{Halpern2019}.

For a consistent, energy preserving discretization, one must retain the anti-symmetry of the right-hand-side operator of Eq.~\ref{eq:AntiDiff} in continuous space. One way to achieve this is by expressing Eq.~\ref{eq:AntiDiff} using the following discrete operators, written as block matrices:
\begin{align}
    \left(\nabla\cdot\vec{v}_\mathbf{K}+\vec{v}_\mathbf{K}\cdot\nabla\right) & \approx \left(\ddiv \mat{V}_{\mathbf{K}}\mathcal{I}+\mathcal{I}^T\mat{V}_{\mathbf{K}}^T\dgrad\right),\label{eq:AntiDiffDisc}\\
    \mat{V}_{\mathbf{K}} & = \begin{bmatrix}
        \mat{v}_{\mathbf{K},x} & 0 & 0\\
        0 & \mat{v}_{\mathbf{K},y} &0 \\
        0 & 0 & \mat{v}_{\mathbf{K},z}
    \end{bmatrix},\\
    \mathcal{I}&=\begin{bmatrix}[1.3]
        \mathcal{I}_x \\
        \mathcal{I}_y \\
        \mathcal{I}_z
    \end{bmatrix},\\
    \mat{v}_\mathbf{K} & \approx 2\mathbf{K}\cdot\dgrad (\log \sqf).
\end{align}Here $\dgrad$ represents the gradient operator, $\ddiv=-\dgrad^T$ represents the divergence operator, and $\mat{V}_\mathbf{K}$ represents the velocity $\vec{v}_\mathbf{K}$ projected as a diagonal matrix. The identity-like column operator $\mathcal{I}$ is used to project scalar quantities to staggered grids as needed for scalar-vector multiplication. The components $\mathcal{I}_{\{x,y,z\}}$ are themselves interpolating matrices, which are symmetric. The operator $\mathcal{I}^T$ collapses vectors quantities into a single scalar. Since we defined $\ddiv=-\dgrad^T$ and $\mat{V}_\mathbf{K}$ is diagonal, the anti-symmetry requirement is satisfied by requiring $\mathcal{I}_x=\mathcal{I}_x^T$, $\mathcal{I}_y=\mathcal{I}_y^T$, and $\mathcal{I}_z=\mathcal{I}_z^T$. The matrix block $\mat{K}$ is the discrete approximation to the diffusion tensor $\tensor{K}$.

Anticipating the analysis of the parallel diffusion problem where $\tensor{K}\sim\bhat\bhat^T$ is a symmetric tensor ($\bhat = [b_x~b_y~b_z]^T$ is the unit magnetic field vector), we introduce the parallel gradient operator
\begin{align}
    \dgrad_\parallel & = \mathcal{B} \mathcal{P}_F \mathcal{B}^T \dgrad.\label{eq:ParGradient}
\end{align}In the expression above $\mathcal{P}_F$ is a square, symmetric operator interpolating the result of the parallel gradient operation as required by grid staggering, and $\mathcal{B}$ contains $b_x$, $b_y$, and $b_z$ along its diagonal. Using the definition above, the parallel Laplacian is $\dgrad_\parallel^2 = \ddiv \mathcal{B} \mathcal{P}_F \mathcal{B}^T \dgrad$, a symmetric operator.

\subsubsection{Finite Volume discretization}
One possible way to discretize Eq.~\ref{eq:AntiDiffDisc} is to use a finite-volume-like approach (FV), where we express fluxes and gradients at the cell boundaries, and project their divergence back at the cell center to obtain the final result. In that case, we define the discrete operators as
\begin{align}
    %
    %
    \dgrad & = \begin{bmatrix}[1.3]
        \dpd{x}{+} \\ \dpd{y}{+} \\ \dpd{z}{+}
    \end{bmatrix},& \text{Gradient}\label{eq:GradFV}\\
    %
    %
    \ddiv & = \begin{bmatrix}
        \dpd{x}{-} & \dpd{y}{-} & \dpd{z}{-} 
    \end{bmatrix} = -\dgrad^T,& \text{Divergence}\\
    %
    %
    \mathcal{I} & = \begin{bmatrix}[1.3]
        \dint{x}{+} \\ \dint{y}{+} \\ \dint{z}{+}
    \end{bmatrix}, & \text{Scalar to vector identity map}\\
    \mathcal{I}^T & = \begin{bmatrix}
        \dint{x}{-} & \dint{y}{-} & \dint{z}{-}
    \end{bmatrix}, & \text{Vector to scalar identity map}\\
    \mathcal{P}_f & = \begin{bmatrix}[1.3]
        1 & \dint{x}{+} \dint{y}{-} & \dint{x}{+} \dint{z}{-} \\
        \dint{x}{-} \dint{y}{+} &1 & \dint{y}{+} \dint{z}{-} \\
        \dint{x}{-} \dint{z}{+} & \dint{y}{-}\dint{z}{+} & 1
    \end{bmatrix}. & \text{Flux projection}\label{eq:VecIdFV}
\end{align}The operators $\dpd{\{x,y,z\}}{\pm}$ are understood to be partial derivatives, with superscripts $(^+,^-)$ indicating mappings back and forth from cell edges. The operators $\dint{\{x,y,z\}}{\pm}$ represent interpolations between grids. These partial derivatives and interpolations are built through the usual Kronecker products of standard finite difference operators, provided in \ref{sec:Stencils} for completeness. As an example, consider the simplest derivative and interpolation operators,
\begin{align}
    \dpd{x}{+} f &= \frac{f_{i+1}-f_i}{\Delta x},~~\dpd{x}{-} = \frac{f_{i-1}-f_i}{\Delta x} = -\left(\dpd{x}{+}\right)^T f,\\
    \dint{x}{+} f &= \frac{f_{i+1}+f_i}{2},~~\dint{x}{-} = \frac{f_{i-1}+f_i}{2} = \left(\dint{x}{+}\right)^T f.
\end{align}Since $\dpd{\{x,y,z\}}{+}= -(\dpd{\{x,y,z\}}{-})^T$ and $\dint{\{x,y,z\}}{+}$ = $(\dint{\{x,y,z\}}{-})^T$ along each direction $x,y,z$, complete anti-symmetry of Eq.~\ref{eq:AntiDiffDisc} can be readily verified.

\subsubsection{Support Operator discretization}
The second approach borrows heavily from the support operator method (SO)~\cite{Shashkov1996}, which was used in \cite{Guenter2005,Guenter2007} to simulate anisotropic heat diffusion in fusion plasmas. Here, we use two diagonally displaced staggered grids, with scalar quantities defined in one grid, and all vector components defined in the other one. In this approach, the operators are defined as
\begin{align}
    %
    %
    \dgrad & = \begin{bmatrix}[1.3]
        \dpd{x}{+} \dint{y}{+} \dint{z}{+} \\ 
        \dint{x}{+} \dpd{y}{+} \dint{z}{+} \\ 
        \dint{x}{+} \dint{y}{+} \dpd{z}{+}
    \end{bmatrix},& \text{Gradient}\label{eq:GradSOM}\\
    %
    %
    \ddiv & = \begin{bmatrix}
        \dpd{x}{-} \dint{y}{-} \dint{z}{-} &
        \dint{x}{-} \dpd{y}{-} \dint{z}{-} & 
        \dint{x}{-} \dint{y}{-} \dpd{z}{-} 
    \end{bmatrix}=-\dgrad^{T},& \text{Divergence}\\
    %
    %
    \mathcal{I} & = \begin{bmatrix}[1.3]
        \dint{x}{+} \dint{y}{+} \dint{z}{+} \\
        \dint{x}{+} \dint{y}{+} \dint{z}{+} \\ 
        \dint{x}{+} \dint{y}{+} \dint{z}{+}
    \end{bmatrix}, & \text{Scalar to vector identity map}\\
    \mathcal{I}^T & = \begin{bmatrix}
        \dint{x}{-} \dint{y}{-} \dint{z}{-} &
        \dint{x}{-} \dint{y}{-} \dint{z}{-} & 
        \dint{x}{-} \dint{y}{-} \dint{z}{-} 
    \end{bmatrix}, & \text{Vector to scalar identity map}\\
    \mathcal{P}_f & = \begin{bmatrix}[1.3]
        1 & 1 & 1 \\
        1 & 1 & 1 \\
        1 & 1 & 1
    \end{bmatrix}. & \text{Flux projection}\label{eq:VecIdSOM}
\end{align}While both of these discretizations (Eqs.~\ref{eq:GradFV}--\ref{eq:VecIdFV} and Eqs.~\ref{eq:GradSOM}--\ref{eq:VecIdSOM}) satisfy the anti-symmetry of Eq.~\ref{eq:AntiDiffDisc}, the use of additional interpolations in Eqs.~\ref{eq:GradSOM}--\ref{eq:VecIdSOM} can have strong implications on the solution quality -- in a counter-intuitive way. This is also true when the standard Laplace operator $\dgrad^2 = \ddiv \dgrad$ is discretized using those definitions, for example, as in \cite{Guenter2005, Guenter2007}. The grid-to-grid mappings have an unexpected influence on the spectral properties of the discrete operators. This topic is investigated below.

\section{Parallel diffusion operator response in wave number space}
\label{sec:Response}
Finite difference approximations to continuous operators result in a truncated wave number space response that resembles, but does not exactly match, the spectral characteristics of the original continuous space operator. The purpose of this section is to develop a qualitative understanding about the wave number-space behavior of the new operators, \emph{i.e.} how good are they at mimicking diffusion? In addition, we are able to compare and contrast the new operators against conventional methods. 

In what follows, the numerical wave number generated by different diffusion operators are obtained. The procedure is standard, and quite similar to the usual Von Neumann stability analysis. We use the name \emph{spectral response function}, and compute it from
\begin{align}
   \mathcal{R}(k) = \left| f^{-1} \mathcal{F}(f) \right|,
\end{align}where we compute the result of $\mathcal{F}$ acting on a plane wave $f=\exp(\imath k x)$, and then remove a factor of $f$. Since all the $\mathcal{R}(k)$'s utilized in this work are real, we report their absolute value. In this context, it will allow us to investigate the numerical wave number produced by different discrete versions of Eq.~\ref{eq:AntiDiff} as well as its conventional counterparts. Note that in the anti-symmetry expressions the calculations involve $\sqf=\exp(\imath k x/2)$ instead of $f=\exp(\imath kx)$, and the half wave number should be easier to resolve at fixed resolution. This suggests that the anti-symmetry operators could result in an improved spectral response with respect to conventional methods. 

In the next two subsections, we progressively analyze 1D and 2D transfer functions, which allows us to (a) develop a qualitative understanding of the anti-symmetry operator spectral response, (b) to compare them against conventional finite difference methods, and then (c) to understand the role of the discrete spectral properties in generating artificial diffusion.

\subsection{One-dimensional spectral transfer}
To estimate the behavior of the new operator in wave number space, it is useful to consider a simplified, 1D diffusion operator based on Eq.~\ref{eq:AntiDiff}. Choosing a trivial diffusion tensor $\tensor{K}=\xhat\xhat^T$ yields $\vec{v}_\mathbf{K} =2\xhat\partial_x \log(\sqf)$, and thus
\begin{align}
     \sqf\left(\nabla\cdot\vec{v}_\mathbf{K}+\vec{v}_\mathbf{K}\cdot\nabla\right)\sqf =
     \sqf\left(\partial_x v_{Kx} + v_{Kx}\partial_x \right)\sqf.\label{eq:AntiDiff1D}
\end{align}Plane wave analysis of Eq.~\ref{eq:AntiDiff1D} in continuous space yields the expected response function for diffusion, $\mathcal{R}(k)=k^2$. We gain additional insight into the numerical behavior of Eq.~\ref{eq:AntiDiff1D} by introducing numerical operators to carry out the derivatives and analyzing the resulting expressions. In 1D, Eqs.~\ref{eq:GradFV}--\ref{eq:VecIdFV} and Eqs.~\ref{eq:GradSOM}--\ref{eq:VecIdSOM} yield the same result:
\begin{align}
    \sqf\left(\partial_x v_{Kx} + v_{Kx}\partial_x \right)\sqf \approx
    \sqf\left(\dpd{x}{-} \mat{V}_{\mathbf{K},x} \dint{x}{+} + \dint{x}{-} \mat{V}^T_{\mathbf{K},x} \dpd{x}{+} \right)\sqf.\label{eq:AntiDiff1DNum}
\end{align}
\begin{figure}
    \centering
    \includegraphics[width=0.49\linewidth]{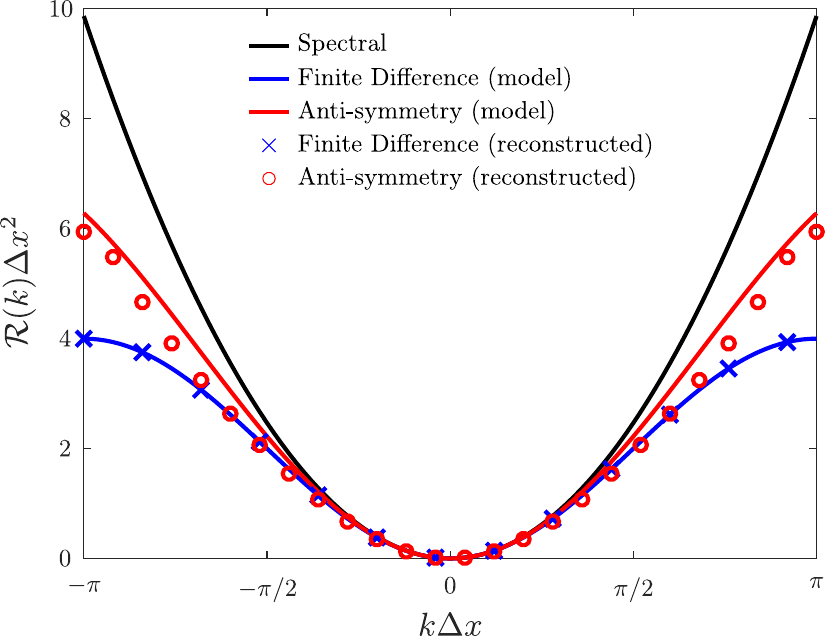}
    \caption{Spectral transfer functions for conventional (blue line) and anti-symmetry (red line) diffusion operators calculated from plane wave analysis using 2nd order accurate finite differences (Eqs.~\ref{eq:DispersionAS} and \ref{eq:DispersionFD}). Crosses and circles of corresponding colors show reconstructed spectral responses obtained from single-step simulations of a diffusing Gaussian pulse. Equation~\ref{eq:DispersionFD} matches its corresponding reconstructed transfer function exactly.}
    \label{fig:Dispersion1D}
\end{figure}Using 2nd order accurate finite difference operators, plane wave analysis yields
\begin{align}
    \mathcal{R}(k) = &\frac{2}{\Delta x^2}\left[2 \left(\frac{k\Delta x}{2}\right) \sin\left(\frac{k\Delta x}{2}\right)\right], &\text{(Anti-symmetry)}\label{eq:DispersionAS}
\end{align}where $\Delta x$ represents the grid spacing. The factor of $k\Delta x / 2$ originates from approximating $\dpd{x}{+} \log [\exp(\imath k x/2)] \approx \imath k/2 $. This means that the $\log$ term should result in a higher-fidelity spectral response than a standard finite difference. Repeating the same analysis on the standard diffusion operator $\partial_{xx}^2 f\approx\dpd{x}{-}\dpd{x}{+} f$ yields
\begin{align}
    \mathcal{R}(k) = &\frac{2}{\Delta x^2}\left[ 1 - \cos\left( k\Delta x\right)\right]. &\text{(Standard finite differences)}\label{eq:DispersionFD}
\end{align}Another remark is useful here: while Eq.~\ref{eq:DispersionFD} uses the normalized wave number $k\Delta x$, only half its value, $k\Delta x/2$, appears in Eq.~\ref{eq:DispersionAS}. This again indicates that Eq.~\ref{eq:DispersionAS} should have better dispersion properties than its conventional counterpart. 

In order to substantiate this claim, the two spectral transfer functions are plotted in Fig.~\ref{fig:Dispersion1D}, and compared against the exact spectral result $\mathcal{R}(k)=k^2$ for $-\pi \leq k \leq \pi$. The highest and lowest wave numbers, $\pi$ and $-\pi$, correspond to the Nyquist cutoff. The conventional diffusion operator (blue line) deviates significantly from the exact value $\mathcal{R}(k)=k^2$ (black line), \emph{i.e.} it is not diffusive enough at high wave number. The anti-symmetry operator (red line) also deviates from the exact result, although it does so far less. This means that it will yield more accurate results than standard finite differences at an extended wave number range.

In practical situations, $f$ is a mix of many Fourier harmonics and Eq.~\ref{eq:AntiDiff1DNum} is non-linear, hence Eq.~\ref{eq:DispersionAS} can become inaccurate. To estimate this effect, we have reconstructed numerically the spectral response of Eq.~\ref{eq:AntiDiff1DNum}. This is carried out by applying the discretized operator onto a narrow Gaussian pulse in a finite difference code, and then extracting the spectral response from the Fourier transform of the result. The reconstructed transfer function (red circles on Fig.~\ref{fig:Dispersion1D}) deviates from the plane wave result, but qualitatively matches it. As additional verification, the same calculation is carried out for the conventional diffusion operator, which is linear. In this case, the result (blue crosses, superimposed with blue solid line), matches Eq.~\ref{eq:DispersionFD} exactly. We conclude from this exercise that Eq.~\ref{eq:DispersionAS} is a reasonable qualitative proxy for the dispersion properties of Eq.~\ref{eq:AntiDiff1DNum}, and that numerical implementations based on Eq.~\ref{eq:AntiDiff1DNum} will likely deviate less from the exact dispersion properties at high wave number.

The spectral accuracy effect found in 1D actually \emph{compounds} in 2D, in particular when using a diffusion tensor that is oblique to the Cartesian axes. As it turns out, spectral transfer at high wave number plays a key role in driving what is commonly denominated as ``artificial diffusion.'' This topic is explored next.

\subsection{Two-dimensional spectral transfer}
\begin{figure}
    \centering
    \begin{tabular}{c@{\hskip0mm}c@{\hskip0mm}c@{\hskip0mm}c@{\hskip0mm}c}
    \hspace{8mm}Fourier & \hspace{8mm}ASFV & \hspace{8mm}ASSO & \hspace{8mm}FV & \hspace{8mm}SO \\
    \includegraphics[width=0.19\linewidth]{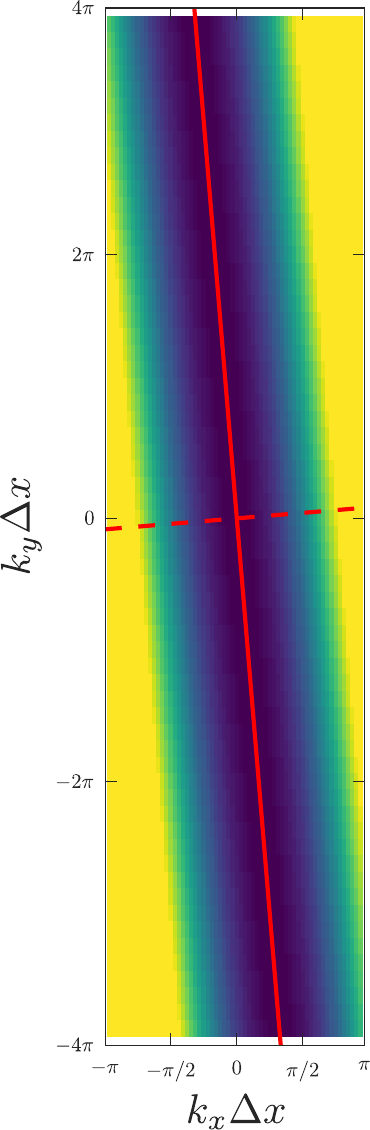}&
    \includegraphics[width=0.19\linewidth]{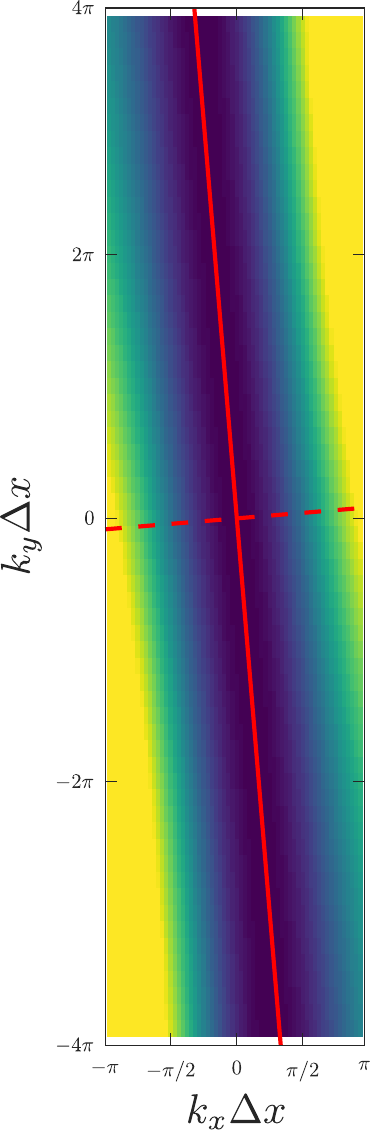}&
    \includegraphics[width=0.19\linewidth]{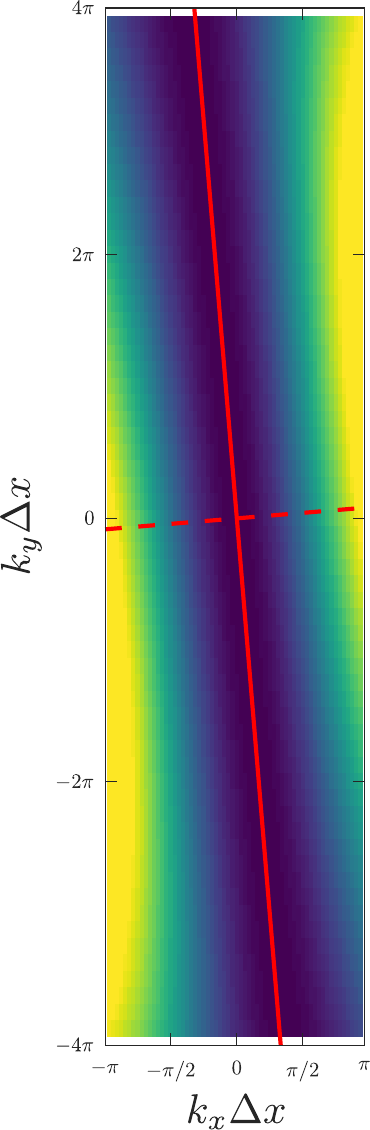}&
    \includegraphics[width=0.19\linewidth]{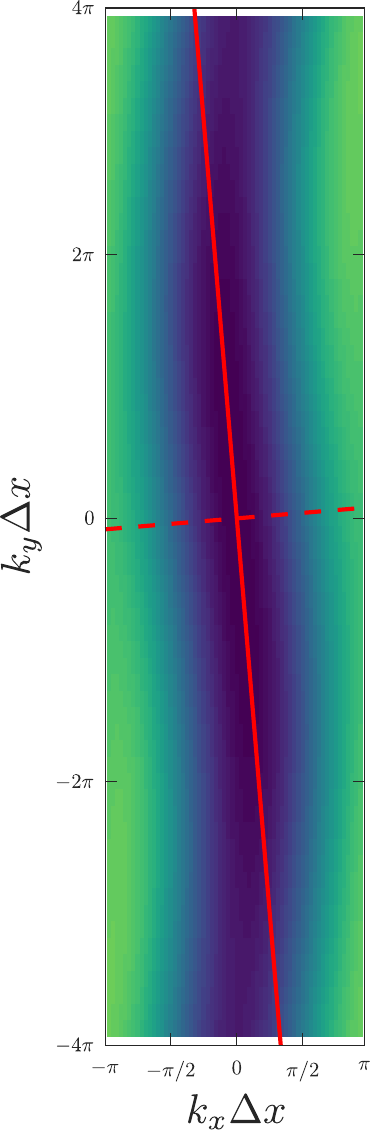}&
    \includegraphics[width=0.19\linewidth]{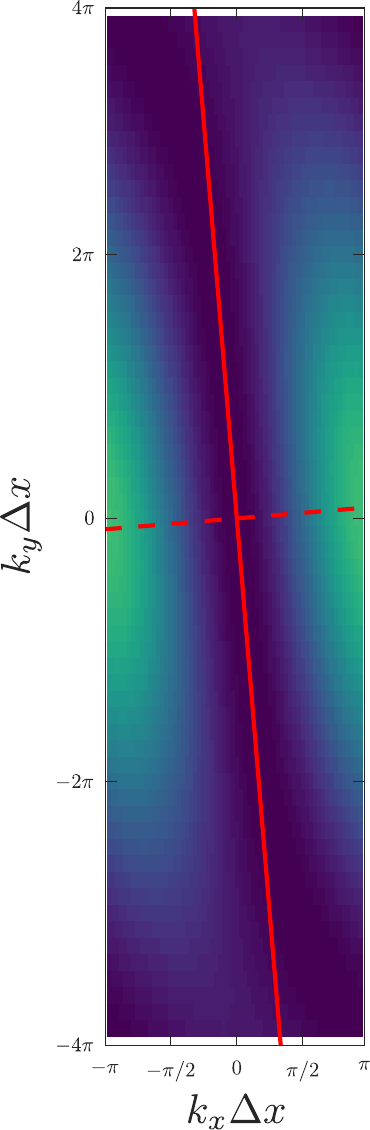} 
    \end{tabular}
    \caption{Discrete spectral response functions for the Fourier, anti-symmetry finite volume (ASFV), anti-symmetry support operator (ASSO), conventional finite volume (FV), and conventional support operator (SO) parallel diffusion operators (Eqs.~\ref{eq:Response2DRef} and \ref{eq:Response2DASFV}--\ref{eq:Response2DSO}, left to right) as a function of the normalized wave numbers $k_x\Delta x$ and $k_y \Delta x$. The red lines indicate $k_\parallel=0$ (solid line) and $k_\perp=0$ (dashed line).}
    \label{fig:Response2D}
\end{figure}

Spectral transfer functions in two dimensions can be calculated in the special case of a diffusion tensor $\tensor{K}=\bhat\bhat^T$ with a constant planar magnetic field. In continuous space, plane wave analysis assuming a plane wave $\sim\exp(-i (k_x x + k_y y) )$ yields
\begin{align}
    \mathcal{R}(k) & = b_x^2 k_x^2 + 2 b_x b_y k_x k_y + b_y^2 k_y^2.\label{eq:Response2DRef}
\end{align}An order of magnitude estimate on the size of each term can be carried out for a field aligned mode. If we approximate $k_y\approx b_x k_x / b_y$, then all of the terms are of the same order of magnitude. To better illustrate Eq.~\ref{eq:Response2DRef}, we use $\bhat=\xhat B_x/B+\yhat B_y/B$, where $B_x = 1$, $B_y = (L_y B_x)/(q L_x)$, $ B^2=B_x^2+B_y^2$. The quantity $q$ represents the number of horizontal turns carried out for each vertical turn as $\bhat$ is followed, and is analogous to the safety factor in tokamaks. Figure~\ref{fig:Response2D} (left panel) shows the result of Eq.~\ref{eq:Response2DRef} computed from a system with $L_x=8\pi$, $L_y=2\pi$, $(n_x,n_y)=(64,64)$, $q=3$. Both axes are normalized using the grid spacing $\Delta x=L_x/n_x$, which preserves the aspect ratio of the $k_x,k_y$ space. The color scale is clipped at a third of its maximum. The red lines indicate, respectively, $k_\parallel = 0$ (solid) and $k_\perp=0$ (dashed). At $k_\parallel=0$ the spectral response is 0, \emph{i.e.} field aligned modes do not diffuse. Also, lines parallel to $k_\parallel=0$ have exactly the same value, which means for a given $k_\parallel$ the spectral response is independent of $k_\perp$. It is also observed that the response is symmetric about both red axes.


The discretized anti-symmetric operators are readily expressed using the parallel gradient and divergence from the previous section,
\begin{align}
    \vec{v}_\mathbf{K} & \approx 2\dgrad_\parallel\left(\log \sqf \right),& \\
    \sqf\left(\nabla \cdot \vec{v}_\mathbf{K}+\vec{v}_\mathbf{K}\cdot\nabla\right)\sqf & \approx \sqf \left( \ddiv \mat{V}_\mathbf{K} \mathcal{I} + \mathcal{I}^T\mat{V}_\mathbf{K}^T \dgrad \right)\sqf \label{eq:AntiDiffPar}
\end{align}
We recall that $\mat{V}_\mathbf{K} = \mat{V}_\mathbf{K}^T$ is a square matrix containing the components of $\vec{v}_\mathbf{K}$ along its diagonal. Corresponding FV and SO discretizations follow by introducing Eqs.~\ref{eq:GradFV}--\ref{eq:VecIdFV} and Eqs.~\ref{eq:GradSOM}--\ref{eq:VecIdSOM} into Eq.~\ref{eq:AntiDiffPar}. Unlike the 1D case, the FV and SO methods lead to different discretizations. The anti-symmetry 2D discrete operators for arbitrary $b_x$ and $b_y$ are
\begin{align}
    \sqf \left( \ddiv \mat{V}_\mathbf{K} \mathcal{I} + \mathcal{I}^T\mat{V}_\mathbf{K}^T \dgrad \right)^{FV} \sqf = & \sqf \bigl[
    \dpd{x}{-} \mat{v}_{\mathbf{K},x}^{FV} \mathbf{I}_x^+ +
    \dpd{y}{-} \mat{v}_{\mathbf{K},y}^{FV} \dint{y}{+} +\nonumber\\
    &\dint{x}{-} \mat{v}_{\mathbf{K},x}^{FV} \dpd{x}{+} +
    \dint{y}{-} \mat{v}_{\mathbf{K},y}^{FV} \dpd{y}{+}
    \bigr] \sqf,\\
    \mat{V}_\mathbf{K}^{FV} & = 2 \begin{bmatrix}[1.3] \left(b_{x} \dint{x}{+} \dint{y}{-} b_{y} \dpd{y}{+} + b_{x}^{2} \dpd{x}{+}\right) \log\sqrt{f} & 0\\0 & \left(b_{y} \dint{x}{-} \dint{y}{+} b_{x} \dpd{x}{+} + b_{y}^{2} \dpd{y}{+}\right) \log\sqrt{f}\end{bmatrix}
\end{align}for the finite volume discretization and 
\begin{align}
     \sqf \left( \ddiv \mat{V}_\mathbf{K} \mathcal{I} + \mathcal{I}^T\mat{V}_\mathbf{K}^T \dgrad \right)^{SO}\sqf = & \sqf \bigl[
     \dpd{x}{-} \dint{y}{-} \mat{v}_{\mathbf{K},x}^{SO} \dint{x}{+} \dint{y}{+} +
     \dint{x}{-} \dpd{y}{-} \mat{v}_{\mathbf{K},y}^{SO} \dint{x}{+} \dint{y}{+} + \nonumber\\
     & \dint{x}{-} \dint{y}{-} \mat{v}_{\mathbf{K},x}^{SO} \dpd{x}{+} \dint{y}{+} +
     \dint{x}{-} \dint{y}{-} \mat{v}_{\mathbf{K},y}^{SO} \dint{x}{+} \dpd{y}{+} 
     \bigr] \sqf,\\
    \mat{V}_\mathbf{K}^{SO} & = 2 \begin{bmatrix}[1.3] \left(b_{x} b_{y} \dint{x}{+} \dpd{y}{+} + b_{x}^{2} \dpd{x}{+} \dint{y}{+}\right) \log\sqrt{f} & 0\\0 & \left(b_{y} b_{x} \dpd{x}{+} \dint{y}{+} + b_{y}^{2} \dint{x}{+} \dpd{y}{+}\right) \log\sqrt{f}\end{bmatrix}
\end{align}for the support operator method. The discrete versions of the conventional operators are computed from $\nabla_\parallel^2 f \approx \ddiv \dgrad_\parallel f \equiv \dgrad^2_\parallel f$:
\begin{align}
    \left(\dgrad_{\parallel}^2\right)^{FV} f & = \left[ \dpd{x}{-} b_{x}^{2} \dpd{x}{+} +
    \dpd{y}{-} b_{y}^{2} \dpd{y}{+} +
    \dpd{x}{-} b_{x} \dint{x}{+} \dint{y}{-} b_{y} \dpd{y}{+} +  \dpd{y}{-} b_{y} \dint{x}{-} \dint{y}{+} b_{x} \dpd{x}{+} \right] f, \label{eq:LaplFV}\\
    \left(\dgrad_{\parallel}^2\right)^{SO} f & = \left[ 
    \dpd{x}{-} \dint{y}{-} b_{x}^{2} \dpd{x}{+} \dint{y}{+} + 
    \dint{x}{-} \dpd{y}{-} b_{y}^{2} \dint{x}{+} \dpd{y}{+} +
    \dpd{x}{-} \dint{y}{-} b_{x} b_{y} \dint{x}{+} \dpd{y}{+} + 
    \dint{x}{-} \dpd{y}{-} b_{y} b_{x} \dpd{x}{+} \dint{y}{+} 
    \right] f.\label{eq:LaplSO}
\end{align}Both operators above are symmetric, which can be verified by taking their respective transposes.

Next, we compute the response functions for all four operators using 2nd order finite differences, under the simplifying assumption that $b_x$ and $b_y$ are constants. We carried out identical analyses with higher order operators, leading to the same qualitative results, however, with the spectral response improving as the discretization accuracy order improves. The response functions for the anti-symmetry based operators are
\begin{align}
    \mathcal{R}^{ASFV}(k) = 2\biggl\{&\left[
    \frac{b_{x}^{2}}{\Delta x^2}\left( k_{x} \Delta x \right)\sin{\left(\frac{k_{x}\Delta x }{2} \right)} +
    \frac{b_{x} b_{y}}{\Delta x\Delta y}\left(  k_{y} \Delta y \right)\sin{\left(\frac{k_{x}\Delta x }{2} \right)}\right]
    +\nonumber\\
    &
    \left[\frac{b_{y}^{2}}{\Delta y^2}\left(k_{y}\Delta y\right)\sin{\left(\frac{k_{y} \Delta y }{2} \right)} + 
    \frac{b_{x} b_{y}}{\Delta x \Delta y}\left(k_{x}\Delta x \right)\sin{\left(\frac{k_{y} \Delta y }{2} \right)}\right]
    \biggr\},\label{eq:Response2DASFV}
    \\
    \mathcal{R}^{ASSO}(k) = 2\biggl\{&
    \left[\frac{b_{x}^{2}}{2 \Delta x^2}
    \left( k_{x} \Delta x \right)\sin{\left(\frac{k_{x}\Delta x }{2} \right)} +
    \frac{b_{x} b_{y}}{2 \Delta x\Delta y} 
    \left( k_{y} \Delta y \right)\sin{\left(\frac{k_{x}\Delta x }{2} \right)}\right]
    \left[1+\cos{\left(\frac{k_y\Delta y}{2}\right)}\right]
    +\nonumber\\
    &
    \left[\frac{b_{y}^{2}}{2\Delta y^2}
    \left(k_{y}\Delta y\right)\sin{\left(\frac{k_{y} \Delta y }{2} \right)} + 
    \frac{b_{x} b_{y}}{2\Delta x \Delta y}
    \left(k_{x}\Delta x \right)\sin{\left(\frac{k_{y} \Delta y }{2} \right)}\right]
    \left[1+\cos{\left(\frac{k_x\Delta x}{2}\right)}\right]
    \biggr\}.
\end{align}The two expressions above show features similar to that of Eq.~\ref{eq:DispersionAS}, namely, there are linear factors of the wave numbers multiplied by the sine of half the normalized wave number $k\Delta$. The support operator discretization provided contains an additional factor of $1+\cos{k\Delta/2}$, acting as a low pass filter. The conventional FV and SO methods lead to
\begin{align}
    \mathcal{R}^{FV}(k) = 2 \biggl\{ & 
     \frac{b_{x}^{2}}{\Delta x^{2}} \left[1-\cos{\left(k_x\Delta x \right)} \right] +
     \frac{b_{x} b_{y} }{\Delta x \Delta y} \sin{\left(k_x\Delta x \right)} \sin{\left(k_y\Delta y \right)} +
     \frac{b_{y}^{2}}{\Delta y^{2}} \left[1-\cos{\left(k_y\Delta y \right)}\right] 
    \biggr\},\label{eq:Response2DFV}
     \\
    \mathcal{R}^{SO}(k) = 2 \biggl\{ & 
     \frac{b_{x}^{2}}{2\Delta x^{2}} \left[1-\cos{\left(k_x\Delta x \right)} \right]
     \left[1+\cos{\left(k_y\Delta y \right)}\right] +
     \frac{b_{x} b_{y} }{\Delta x \Delta y} \sin{\left(k_x\Delta x \right)} \sin{\left(k_y\Delta y \right)} + \nonumber\\
     &
     \frac{b_{y}^{2}}{2\Delta y^{2}} \left[1-\cos{\left(k_y\Delta y \right)}\right]
     \left[1+\cos{\left(k_x\Delta x \right)} \right]
    \biggr\}.\label{eq:Response2DSO}
\end{align}The FV method reads as a straightforward 2D generalization of Eq.~\ref{eq:DispersionFD} with a cross derivative term added, while the SO also contains a low-$k$ pass filter $1+\cos{k\Delta}$ in the direction perpendicular to each second derivative.

Figure~\ref{fig:Response2D} shows 2D images of Eqs.~\ref{eq:Response2DASFV}--\ref{eq:Response2DSO} (2nd panel left to right) using the same parameters as used for Eq.~\ref{eq:Response2DRef} before. A dashed red line represents the $\mathbf{k}_\parallel$ axis and a solid red line represents the $\mathbf{k}_\perp$ axis. The $\mathcal{R}(k)$ from the anti-symmetry operators resemble the left panel of Fig.~\ref{fig:Response2D} much better than the conventional operators. However, for all operators, it is observed that (a) symmetry around the $\mathbf{k}_\perp$ and $\mathbf{k}_\parallel$ axes is lost and (b) lines with equal $k_\parallel$ do not have the same color. The asymmetry effect (a) is especially prominent in the SO operator. We notice, in particular, that the $(1+\cos{k})$ factor introduces a significant spectral cutoff at large $k_\perp$ -- modes with large $k_\perp$ diffuse very slowly. Point (b) means that modes with different $k_\perp$ at a fixed $k_\parallel$ will undergo diffusion at different rates. Thus, we have pinpointed the origin of the so-called ``artificial diffusion.'' 
\begin{figure}
    \centering
    \includegraphics[width=0.35\linewidth]{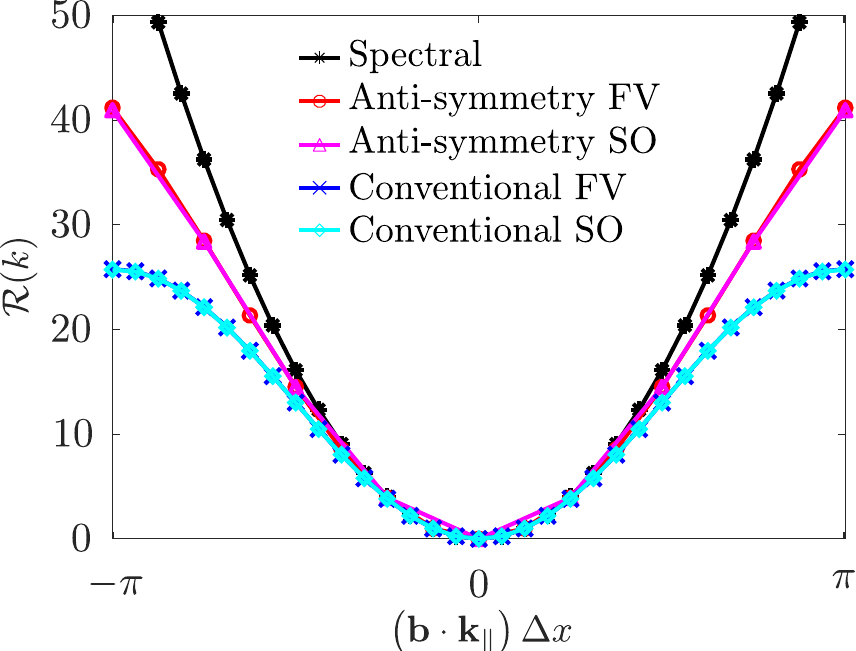}
    \includegraphics[width=0.35\linewidth]{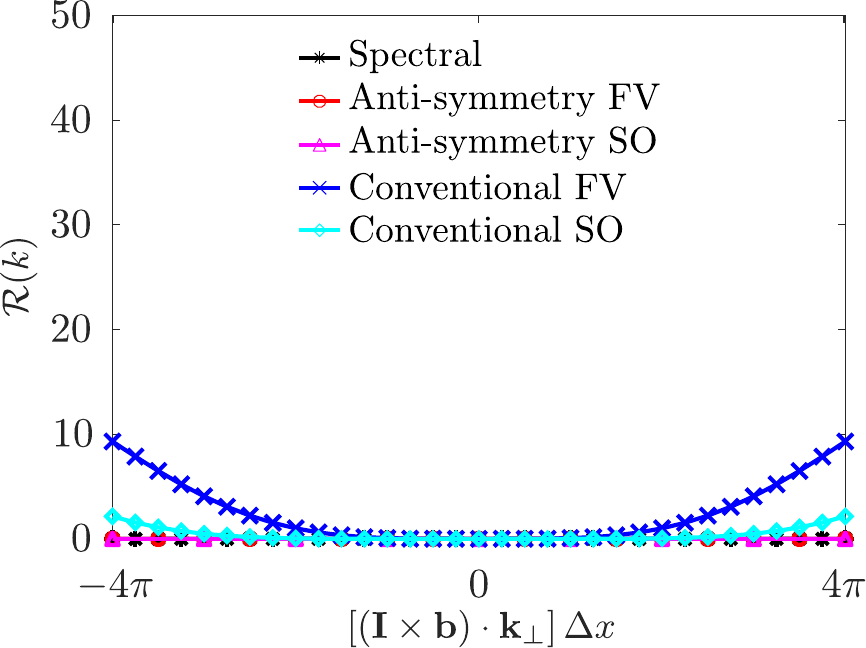}
    \caption{Response functions (Eqs.~\ref{eq:Response2DASFV}--\ref{eq:Response2DSO}) for each method plotted as a function of $k_\parallel$ with $k_\perp=0$ (left panel) and $k_\perp$ with $k_\parallel=0$ (right panel). Some of the curves are superimposed (left panel: red and magenta, blue and cyan; right panel: red, magenta, black)}
    \label{fig:Response2DAxes}
\end{figure}
As further illustration, we plot the response functions as a function of $k_\perp$ and $k_\parallel$ (Fig.~\ref{fig:Response2DAxes} left and right panels). The curves are obtained from 2D interpolation of the response functions, which are defined in $k_x,k_y$ space instead of $k_\perp,k_\parallel$ space because the discretization is not field aligned. The left panel shows the response functions as a function of $k_\parallel$ for $k_\perp=0$. Since the low pass filter has no effect at $k_\perp=0$, the finite volume and support operator responses are superimposed. This result is identical to Fig.~\ref{fig:Dispersion1D} up to a constant. The right panel shows the response functions for $k_\parallel=0$ as a function of $k_\perp$. We observe that modes away from $k_\perp=0$ undergo an artificial, numerical  diffusion process (non-zero response), stemming from the inaccurate spectral properties of the discrete parallel Laplacian operators. The effect is particularly large for the conventional finite volume operator, with the largest values of $\mathcal{R}(k)$ at large $k_\perp$ only a factor of 3 smaller than the largest $\mathcal{R}(k)$ at large $k_\parallel$. The artificial diffusion generated by the SO method is significantly smaller. However, this is not because of improved numerical qualities at high $k_\perp$. Rather, the artificial diffusion is suppressed because of the effect of the low pass filters, which eliminate \emph{most but not all} of the large $k_\perp$ spurious response.


While the spurious behavior at large $k_\perp$ appears to be strongly suppressed for the anti-symmetry based algorithms, we recall that in Fig.~\ref{fig:Dispersion1D} (a simpler 1D case) the anti-symmetry response function was intermediate between Eq.~\ref{eq:DispersionAS} and Eq.~\ref{eq:DispersionFD}. Although we expect a reasonable improvement using anti-symmetry with respect to conventional methods, we do not expect a complete suppression of artificial diffusion. This has motivated a comprehensive test of the 4 algorithms, which allows us to evaluate their spectral accuracy and to test the results of this section. The results of this exercise are described below.

\section{Numerical tests}
\label{sec:Simulations}
In this Section, we carry out simulations to evaluate the performance of the anti-symmetry parallel diffusion algorithms developed in Sec.~\ref{sec:Algorithms} and to compare them to their conventional counterparts. The calculations involve a pure parallel diffusion problem in simple Cartesian geometry, which allows us to evaluate the artificial diffusion incurred by each method. All four algorithms are implemented as an initial value code using the \alma{} numerical engine~\cite{Halpern2021}. The equations evolved in each representation are
\begin{align}
    \vec{v}_\mathbf{K} &= 2\kappa_\parallel\bhat\bhat^T\left(\nabla\log{\sqf}\right), & \nonumber\\
    \ddt{\sqf}{t} &= \frac{1}{2}\left(\nabla\cdot\vec{v}_{\mathbf{K}}+\vec{v}_{\mathbf{K}}\cdot\nabla \right)\sqf, &\text{Anti-symmetry}\\
    \ddt{f}{t} &= \kappa_\parallel\nabla^2_\parallel f, &\text{Conventional}\label{eq:ConvDiff}
\end{align}implemented with the discrete gradient and divergence operators detailed in Section~\ref{sec:Algorithms}. Discretizations are carried out using 2nd, 4th, and 6th order finite difference stencils. A high precision (RK4) time integrator is used in order to avoid time integration error from interfering with error measurement. Simulations of a test case using a high-order discretization were carried out with varying timesteps as a diagnostic, finding essentially no difference in the final result. Whenever possible, the timestep is chosen at the lowest resolution used for each problem, and then adjusted accordingly using the standard Courant-Friedrichs-Lewy (CFL) criterion for diffusion $\Delta t_{max}\propto 1/\Delta x^2$.

Simulations of parallel diffusion are carried out in increasingly more challenging settings. First, we use a 2D, planar, constant magnetic field case where the analytical solution can be computed exactly. This case is used to extensively verify our algorithms and to measure their spectral accuracy, in particular, to measure how much artificial diffusion they generate. Then, still in 2D, we use a circular magnetic field to evaluate the performance of the methods with non-uniform magnetic fields. Finally, we carry out simulations in a screw pinch configuration, mirroring a test case in Stegmeir \emph{et al.}~\cite{Stegmeir2023} that compares the properties of the FCI method and the SO method (GU2 and GU4 methods in their notation).

\begin{figure}
    \centering
    \begin{tabular}{c@{\hskip2mm}c@{\hskip0.5mm}c}
    \rotatebox{90}{\hspace{0.055\linewidth}$y$} & \includegraphics[width=0.48\linewidth]{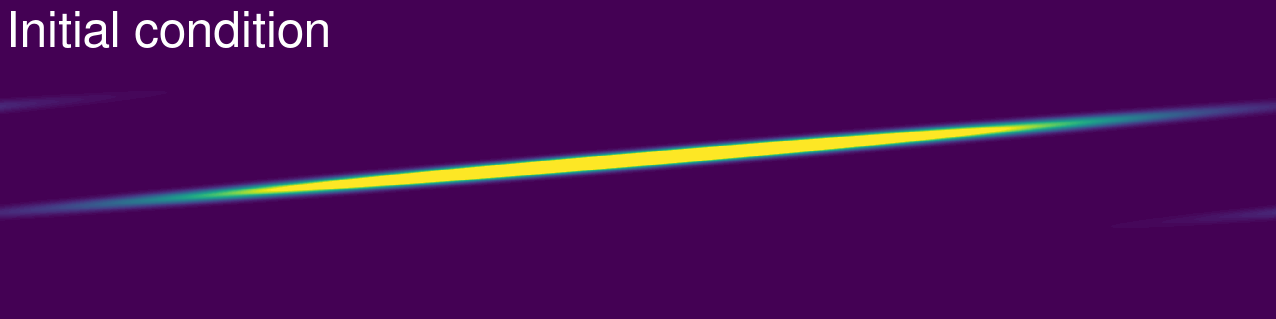}&
    \includegraphics[width=0.48\linewidth]{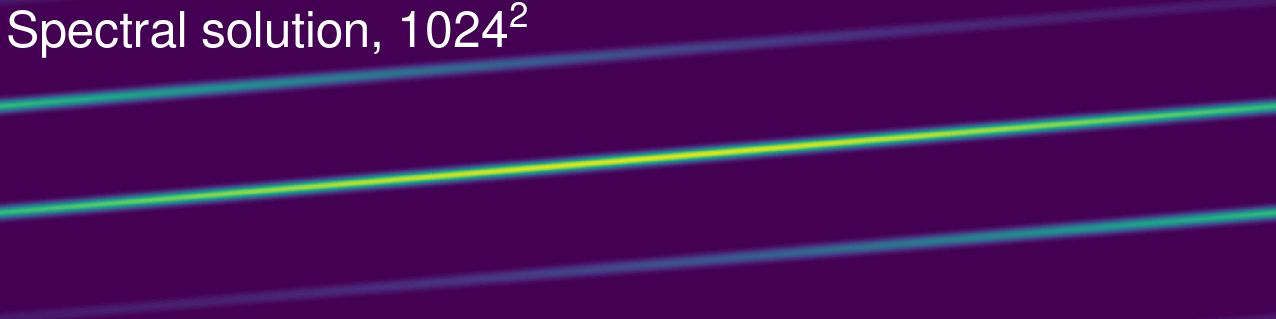}\\
    & $x$ & $x$
    \end{tabular}
    \caption{Left: Field aligned Gaussian (Eq.~\ref{eq:f0}) initialized with parallel and perpendicular widths $\sigma_\parallel=5$, $\sigma_\perp=0.083$, $q=3$ turns around the $x$ axis. Right: Gaussian from left panel at $t=10$ after undergoing parallel diffusion with $\kappa_\parallel=10$ (obtained using Eq~\ref{eq:Solution2D}).}
    \label{fig:InitialCondition}
\end{figure}

\begin{figure}[ht]
    \centering
    \includegraphics[height=0.25\linewidth]{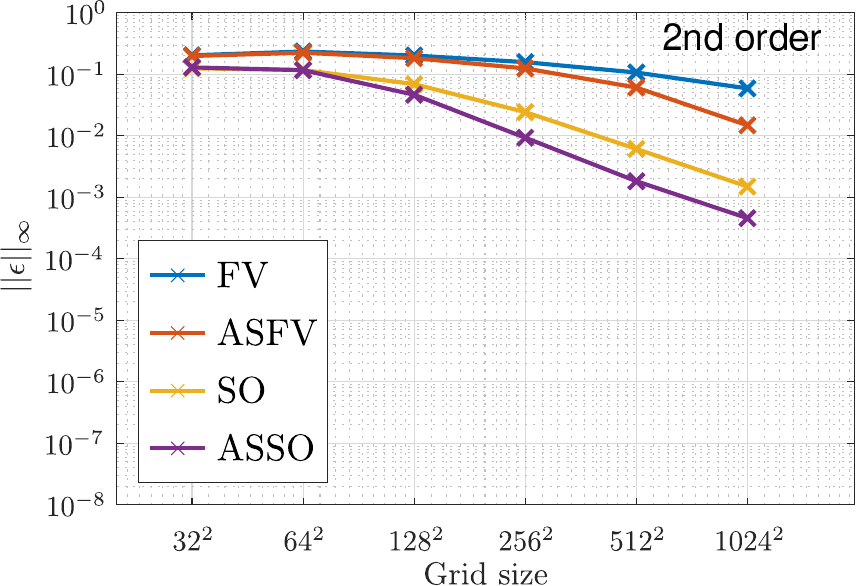}
    \includegraphics[height=0.25\linewidth]{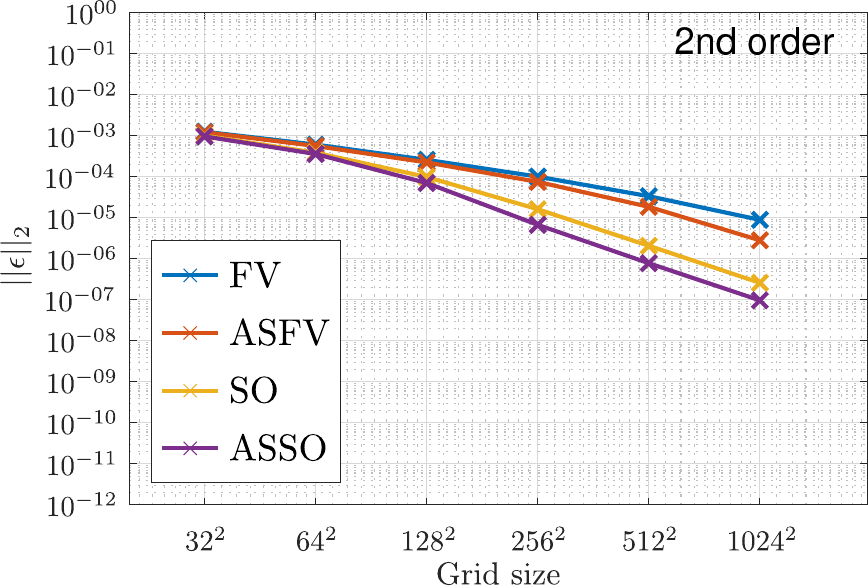}\\    \includegraphics[height=0.25\linewidth]{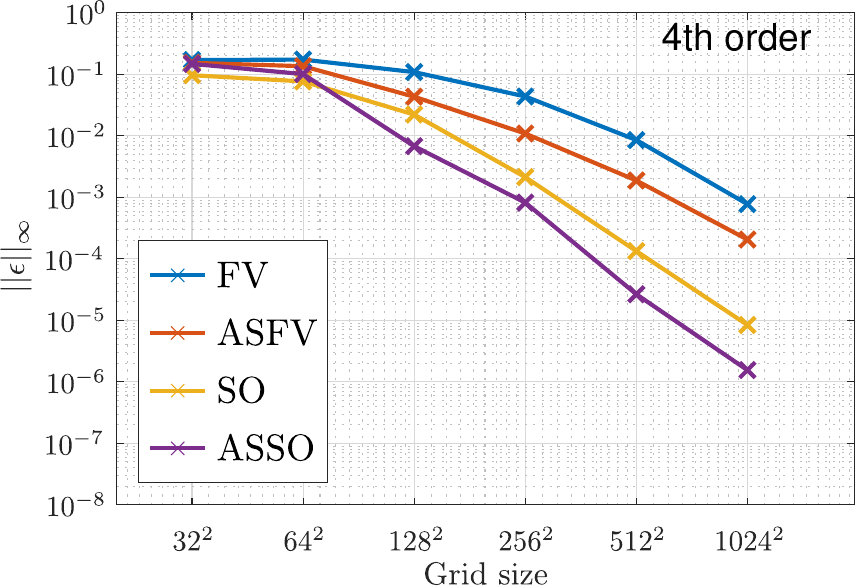}
    \includegraphics[height=0.25\linewidth]{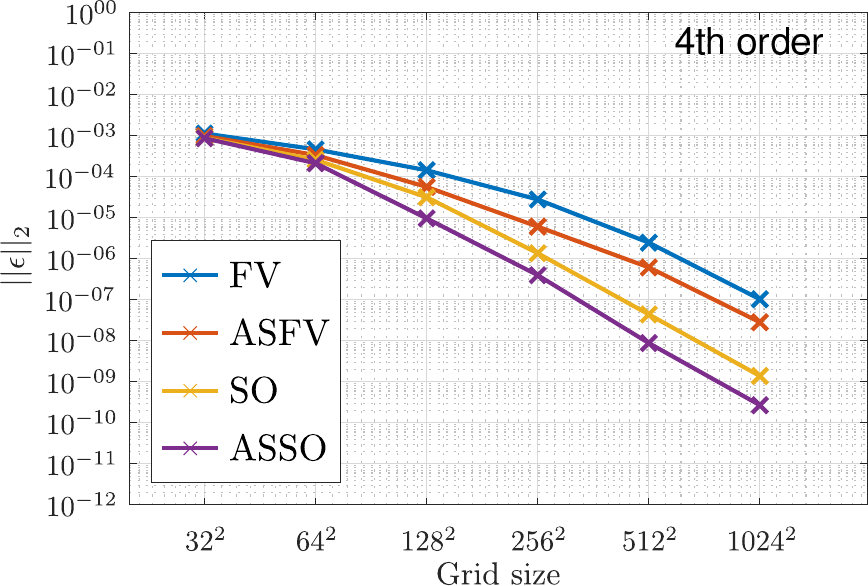}\\
    \includegraphics[height=0.25\linewidth]{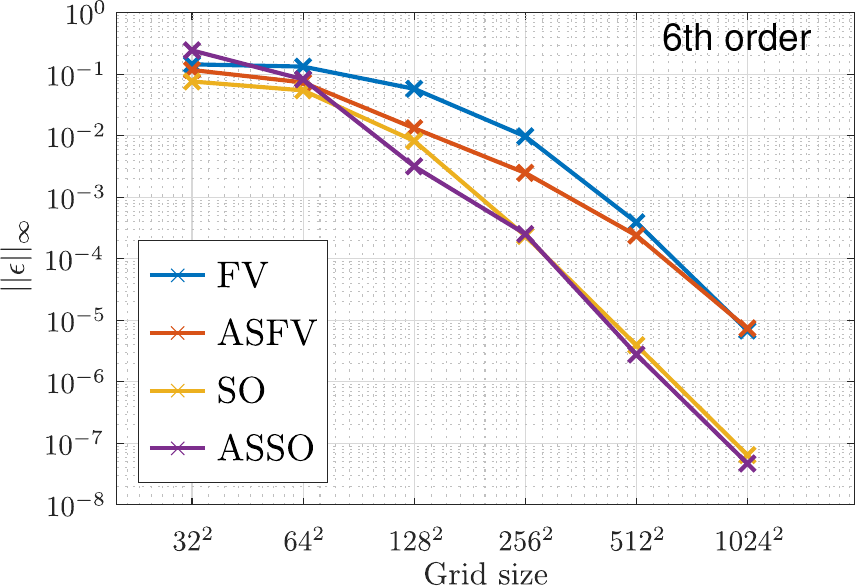}
    \includegraphics[height=0.25\linewidth]{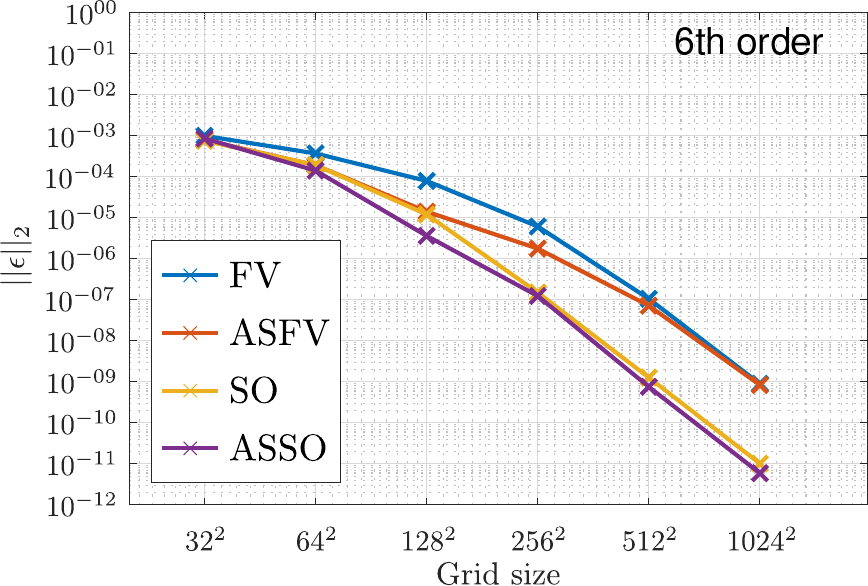}
    \caption{Infinity norm error (left column) and square norm error (right column) in 2D simulations of parallel diffusion in a constant planar magnetic field. Anti-symmetry (AS) and conventional representations are used, implemented using finite volume (FV) and support operator (SO) methods of 2nd, 4th, and 6th order accuracy (top, center, bottom rows).}
    \label{fig:Convergence}
\end{figure}
\subsection{Parallel diffusion in a planar constant magnetic field}\label{sec:Simulations2D}
The simulation setup mirrors the one used in Section~\ref{sec:Response} to compute the response functions. We use a 2D periodic Cartesian domain with $0\leq x<L_x=8\pi$, $0\leq y<L_y=2\pi$, and a constant magnetic field $\vec{B}=B_x\xhat+B_y\yhat$. The spatially uniform magnetic field components are given by $B_x=1$ and $B_y=(L_y B_x)/(L_x q)$. The quantity $q$, resembling the tokamak safety factor, describes the number of horizontal ($x$) field line turns for each vertical ($y$) turn. The parallel conductivity is $\kappa_\parallel=10$, and simulations are evolved to $t=10$. The initial condition $f(t=0)$ is a field-aligned Gaussian
\begin{align}
    f(t=0)&=\exp\left(-\frac{l_\perp^2}{2\sigma_{\perp 0}^2} -\frac{l_\parallel^2}{2\sigma_{\parallel 0}^2}\right) + f_{\mathrm{bg},0}\label{eq:f0}
\end{align}with initial perpendicular and parallel widths $\sigma_{\perp 0}=0.083$ and $\sigma_{\parallel 0}=5$. The parallel and perpendicular coordinates are computed from a solid rotation of the Cartesian coordinates, $l_\parallel = (x-x_0) B_x / B + (y-y_0) B_y / B$ and $l_\perp = -(x-x_0) B_y / B + (y-y_0) B_x / B$. The peak of the Gaussian is placed at $x_0=L_x/2$ and $y_0=L_y/2$, corresponding to the center of the domain. A small constant $f_{\mathrm{bg},0}=10^{-3}$ is added to the initial condition to avoid $\log (0)$ in the anti-symmetry algorithms. 

\begin{sidewaysfigure}[!]
    \centering
    \begin{tabular}{c@{\hskip2mm}c@{\hskip0.5mm}c@{\hskip0.5mm}c}
    \rotatebox{90}{\hspace{0.037\linewidth}$y$} &
    \includegraphics[width=0.32\linewidth]{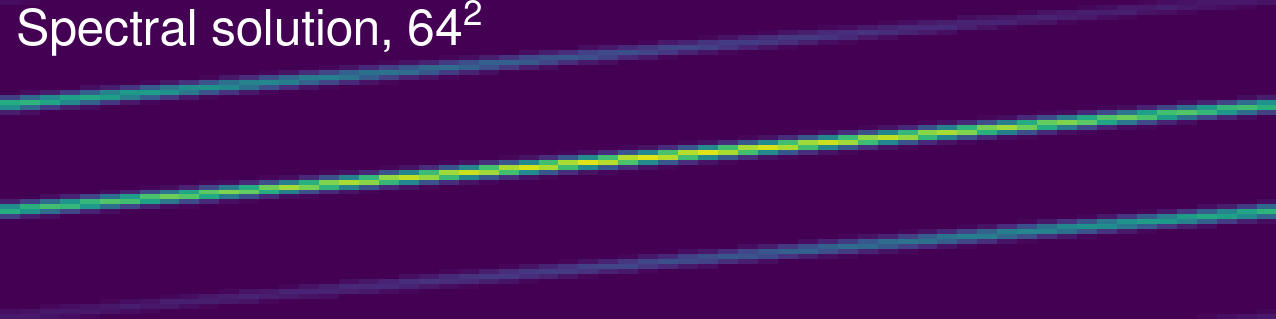} &
    \includegraphics[width=0.32\linewidth]{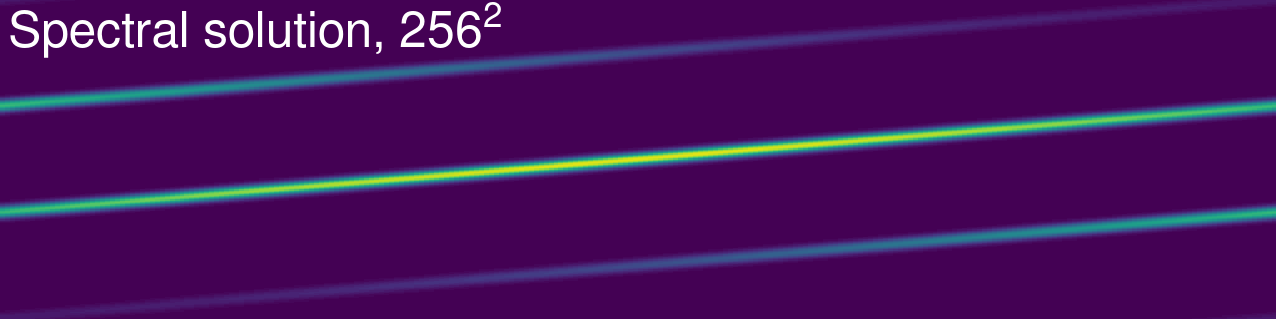} &
    \includegraphics[width=0.32\linewidth]{Figures/fft_x1024.png}\\
    \rotatebox{90}{\hspace{0.037\linewidth}$y$} &
    \includegraphics[width=0.32\linewidth]{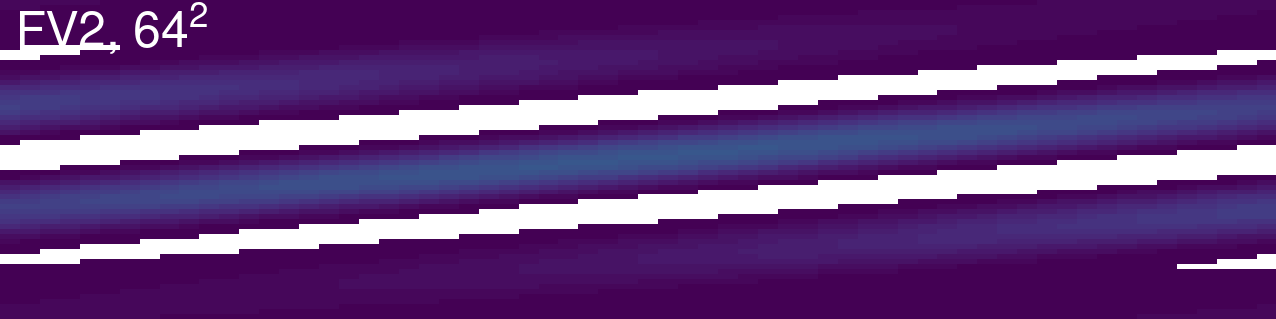} &
    \includegraphics[width=0.32\linewidth]{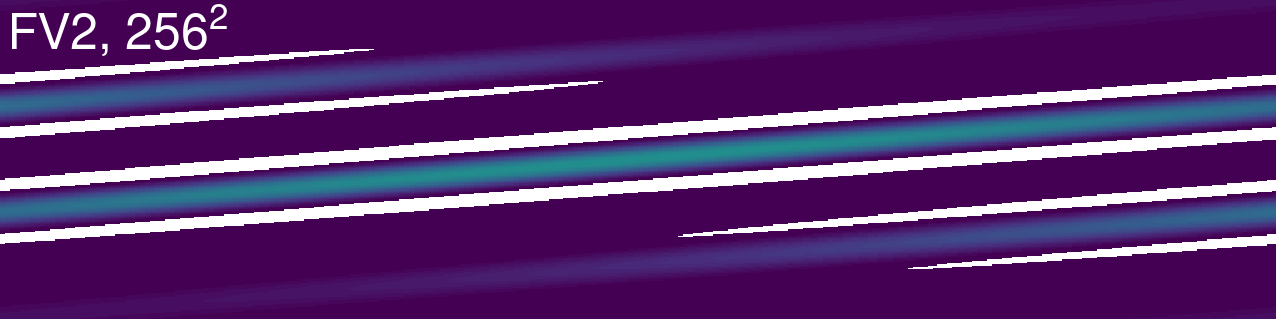} &
    \includegraphics[width=0.32\linewidth]{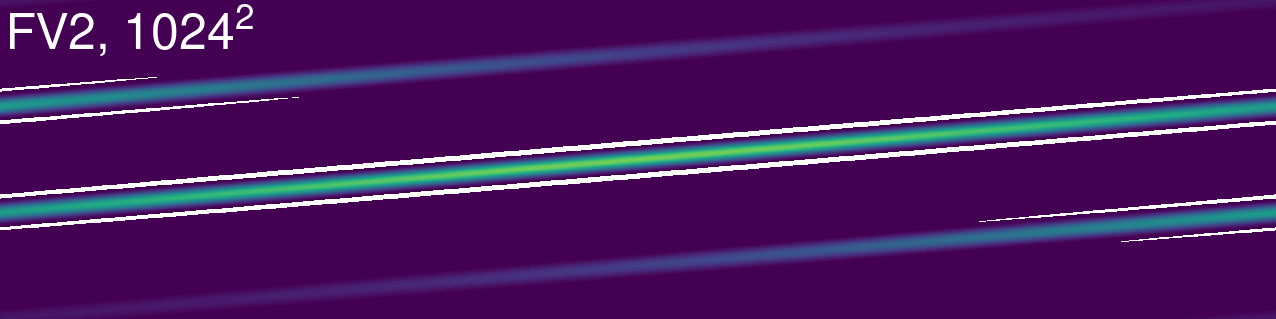}\\
    \rotatebox{90}{\hspace{0.037\linewidth}$y$} &
    \includegraphics[width=0.32\linewidth]{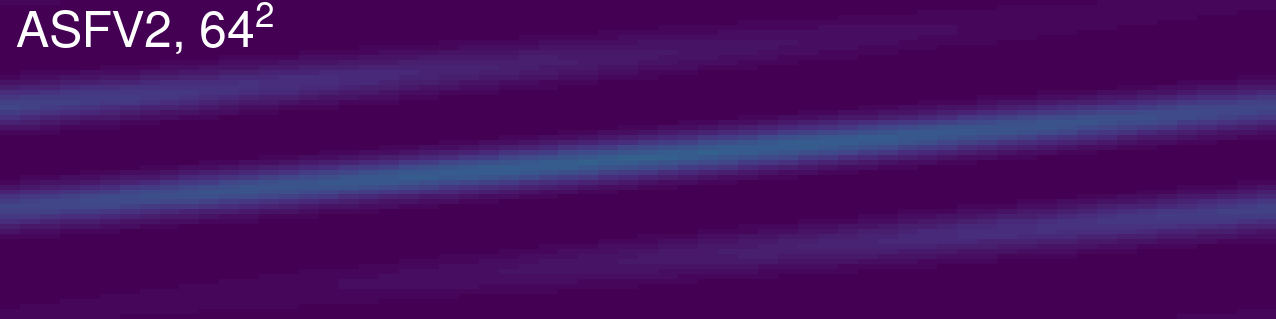} &
    \includegraphics[width=0.32\linewidth]{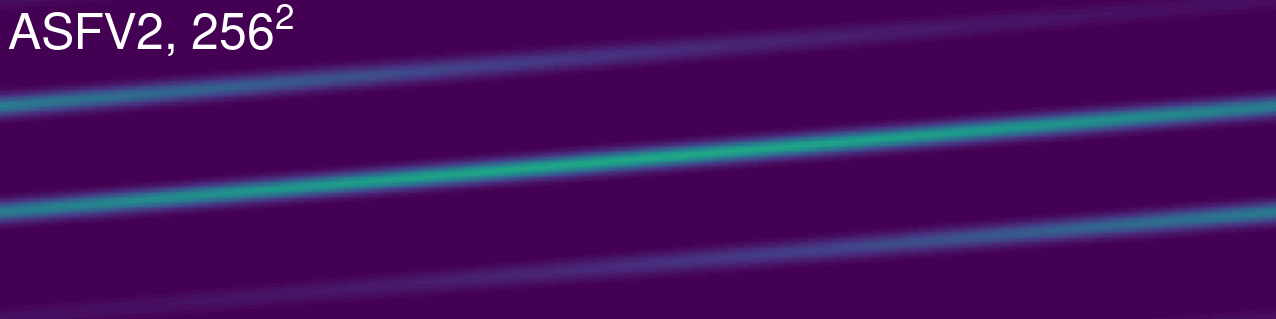} &
    \includegraphics[width=0.32\linewidth]{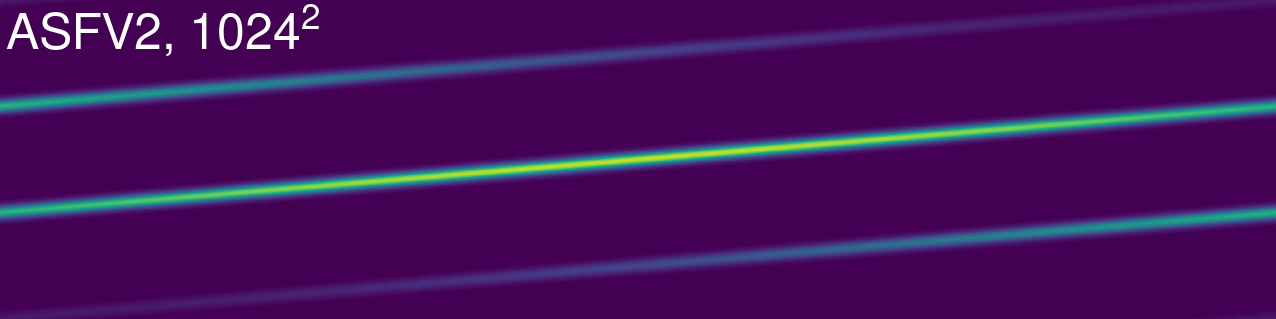}\\
    \rotatebox{90}{\hspace{0.037\linewidth}$y$} &
    \includegraphics[width=0.32\linewidth]{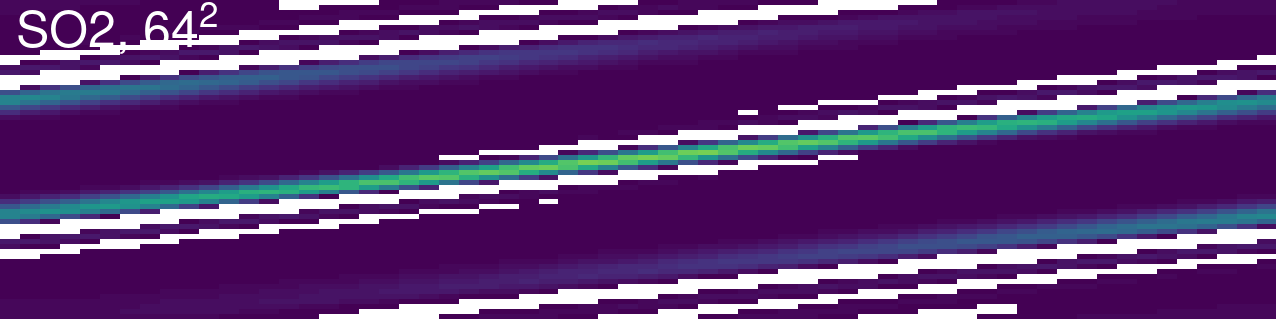} &
    \includegraphics[width=0.32\linewidth]{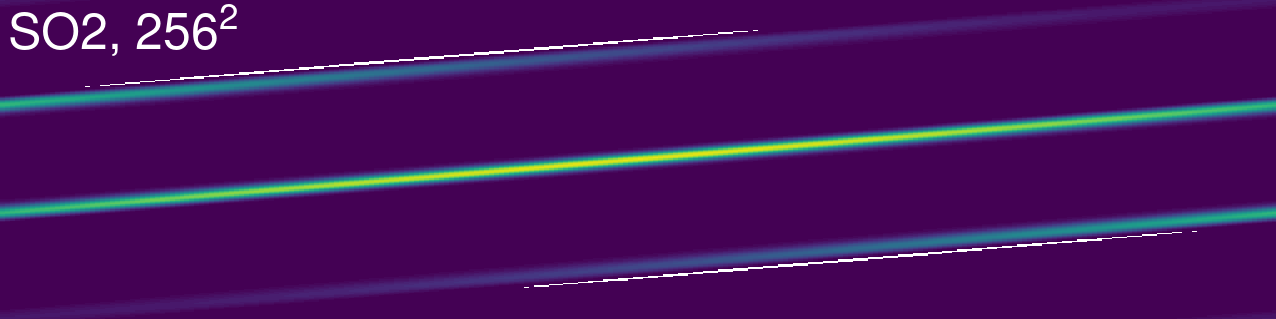} &
    \includegraphics[width=0.32\linewidth]{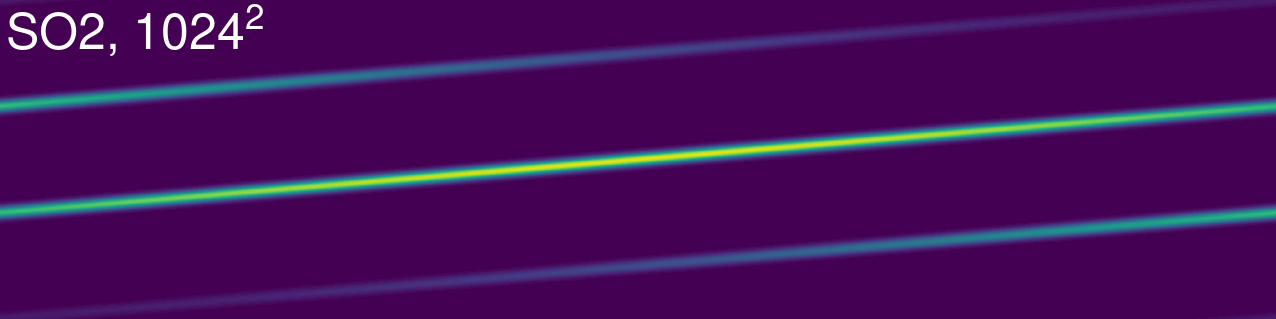}\\
    \rotatebox{90}{\hspace{0.037\linewidth}$y$} &
    \includegraphics[width=0.32\linewidth]{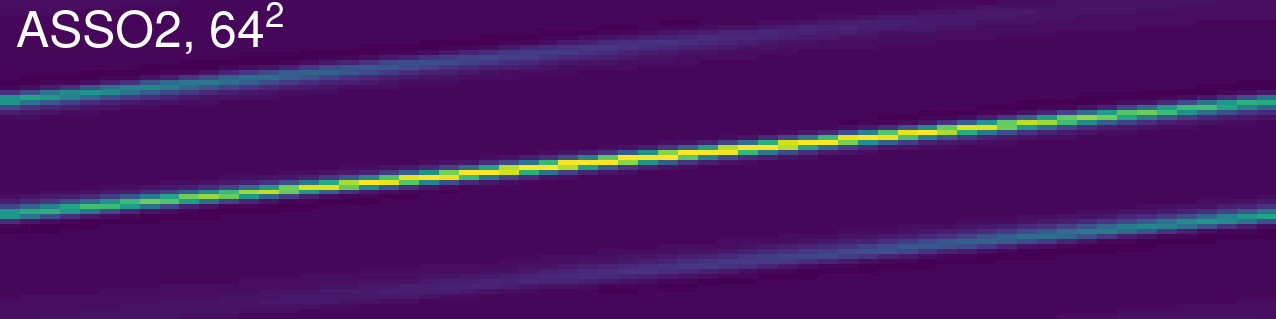} &
    \includegraphics[width=0.32\linewidth]{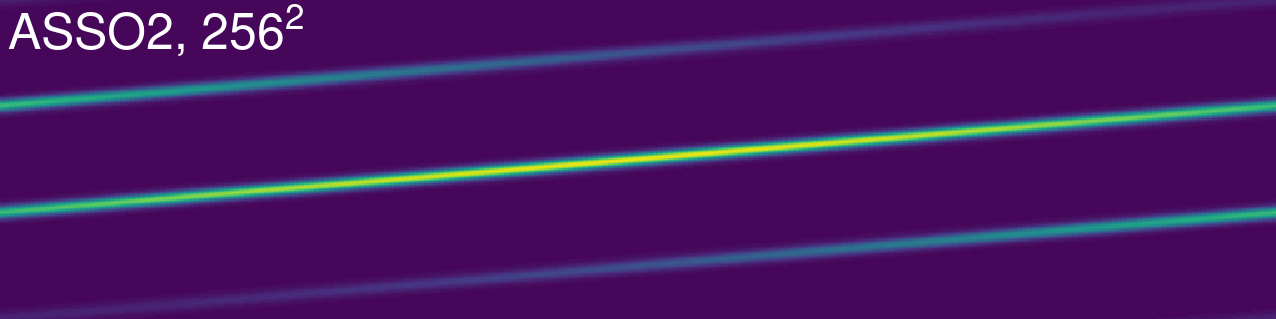} &
    \includegraphics[width=0.32\linewidth]{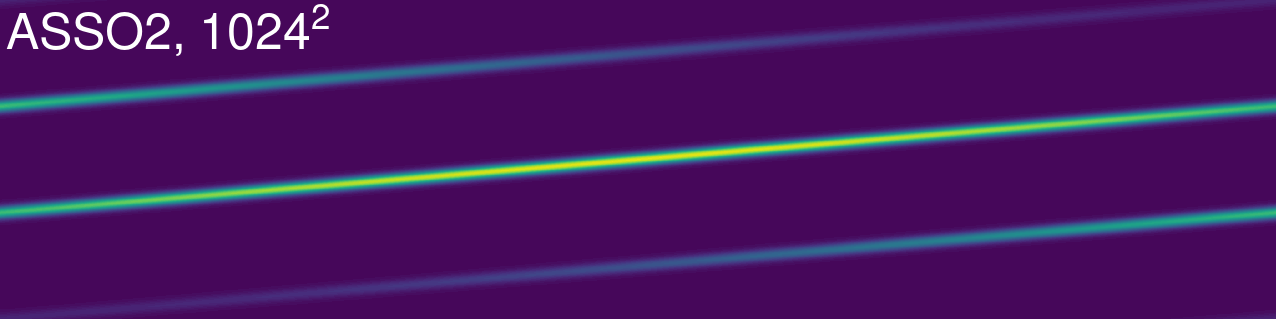} \\
    & $x$ & $x$ & $x$
    \end{tabular}
    \caption{Numerical solution of Eq.~\ref{eq:ConvDiff} with $\kappa_\parallel=10$ at $t=10$ in a planar constant magnetic field. Solutions are computed using Fourier series (Eq.~\ref{eq:Solution2D}) (top row),
    conventional finite volume (second row, FV),
    anti-symmetry finite volume (third row, ASFV),
    conventional support operator (fourth row, SO),
    and
    anti-symmetry support operator (fifth row, ASSO) algorithms, all using 2nd order accurate stencils. Grid resolutions are $(n_x,n_y)=(64,64),(256,256),(1024,1024)$ (left, center, right). Identical color scale is used for all images, with white color indicating positivity violations.}
    \label{fig:profiles1}
\end{sidewaysfigure}

Since the magnetic field is spatially uniform, Equation~\ref{eq:ConvDiff} can be solved using Fourier transforms:
\begin{align}
    f(t) = \mathcal{F}^{-1}\left[\mathcal{F}\left(f(t=0)\right)\exp\left(-k^2_\parallel \kappa_\parallel t\right) \right],\label{eq:Solution2D}
\end{align}where $\mathcal{F}$ and $\mathcal{F}^{-1}$ represent the Fourier transform and its inverse, and $f(t=0)$ is the initial condition. The parallel wave number $k_\parallel$ is given in Eq.~\ref{eq:Response2DRef}.

In simple diffusion processes, the width of a Gaussian pulse can be modeled as $\sigma^2 = 2\kappa t$. This expression can be used to define \emph{effective} parallel and perpendicular conductivities,
\begin{align}
    \kappa_{\parallel \mathrm{eff}}&=\frac{\sigma_\parallel^2 - \sigma_{\parallel 0}^2}{2t},\label{eq:SpreadPar}\\
    \kappa_{\perp \mathrm{eff}}&=\frac{\sigma_\perp^2 - \sigma_{\perp 0}^2}{2t}\label{eq:SpreadPerp}.
\end{align}At the end of our simulations, based on Eqs.~\ref{eq:SpreadPar} and~\ref{eq:SpreadPerp}, we expect $\sigma_\parallel=15$ and $\sigma_\perp=0.083$. Since $\kappa_\perp=0$, any increase in perpendicular width is attributed to artificial numerical diffusion.

As an illustration, we plot the initial condition for a case with $q=3$ in Fig.~\ref{fig:InitialCondition} (left panel), together with its Fourier solution (Eq.~\ref{eq:Solution2D}) at $t=10$ using a grid $(n_x,n_y)=(1024,1024)$ (right panel). There is no perpendicular spread, and the parallel length agrees with that predicted by Eq.~\ref{eq:SpreadPar}. Since Eq.~\ref{eq:Solution2D} is, in practice, a truncated Fourier series, a sufficiently large resolution is needed for a converged solution. It was found that the Fourier series saturates around $(n_x,n_y)=(256,256)$. Therefore, cases below that resolution should be considered as under-resolved. 

\begin{sidewaysfigure}[!]
    \centering
    \begin{tabular}{c@{\hskip2mm}c@{\hskip0.5mm}c@{\hskip0.5mm}c}
    \rotatebox{90}{\hspace{0.037\linewidth}$y$} &
    \includegraphics[width=0.32\linewidth]{Figures/fv2_x64.png} &
    \includegraphics[width=0.32\linewidth]{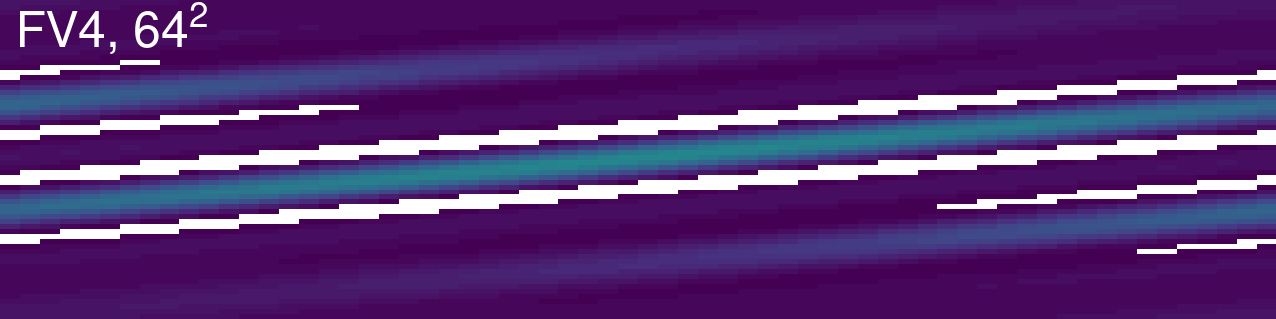} &
    \includegraphics[width=0.32\linewidth]{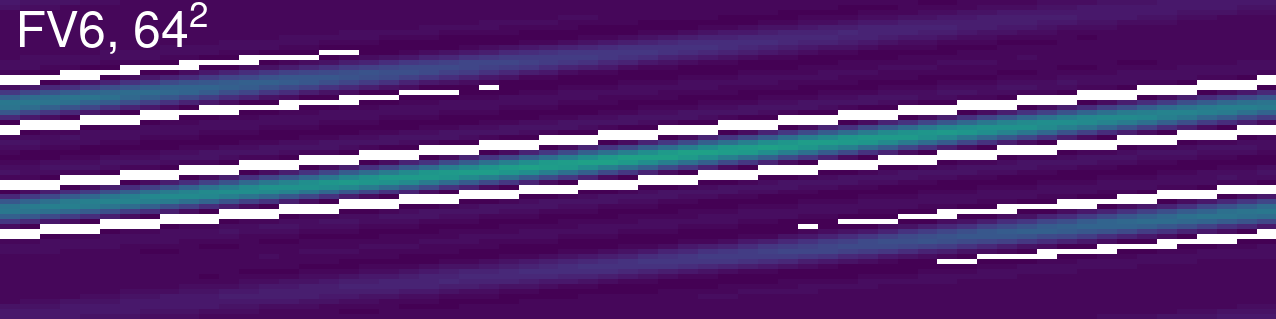}\\
    \rotatebox{90}{\hspace{0.037\linewidth}$y$} &
    \includegraphics[width=0.32\linewidth]{Figures/as2_x64.png} &
    \includegraphics[width=0.32\linewidth]{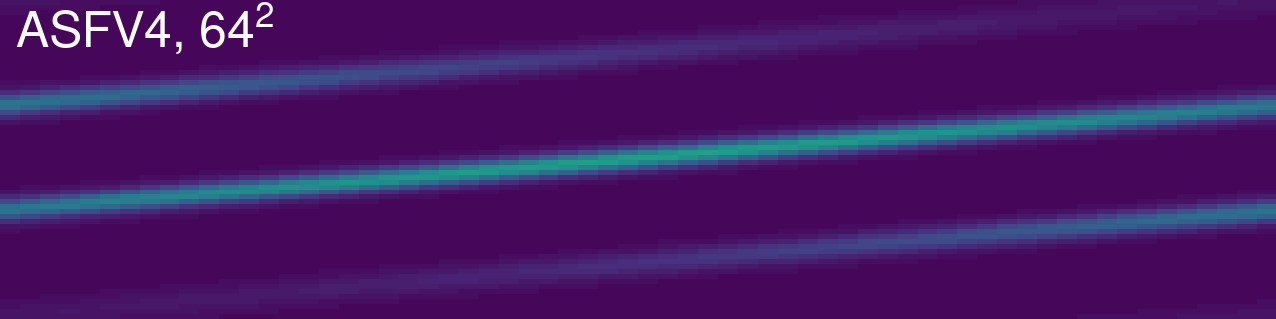} &
    \includegraphics[width=0.32\linewidth]{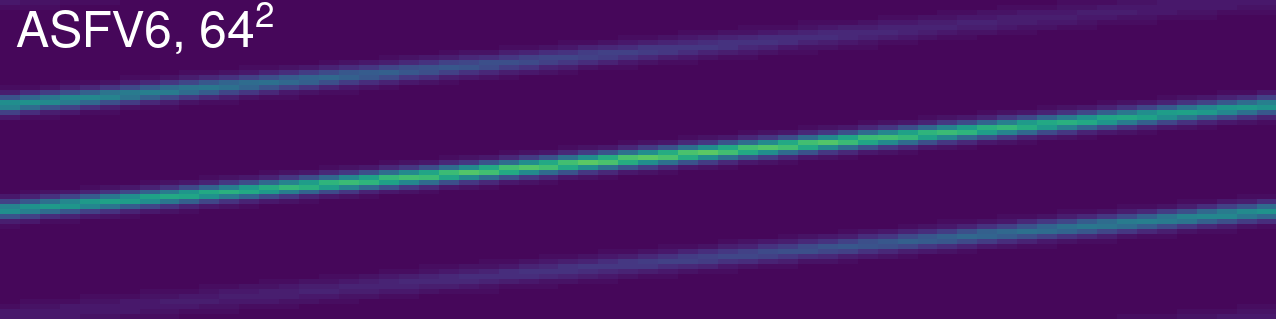}\\
    \rotatebox{90}{\hspace{0.037\linewidth}$y$} &
    \includegraphics[width=0.32\linewidth]{Figures/so2_x64.png} &
    \includegraphics[width=0.32\linewidth]{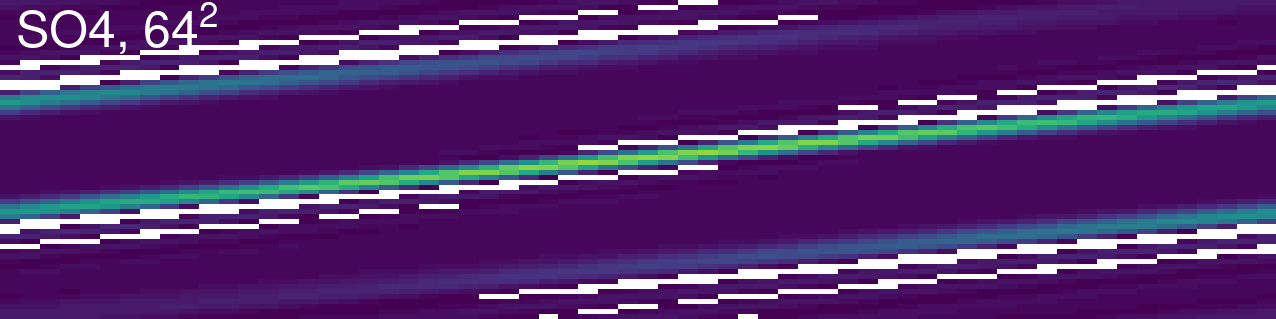} &
    \includegraphics[width=0.32\linewidth]{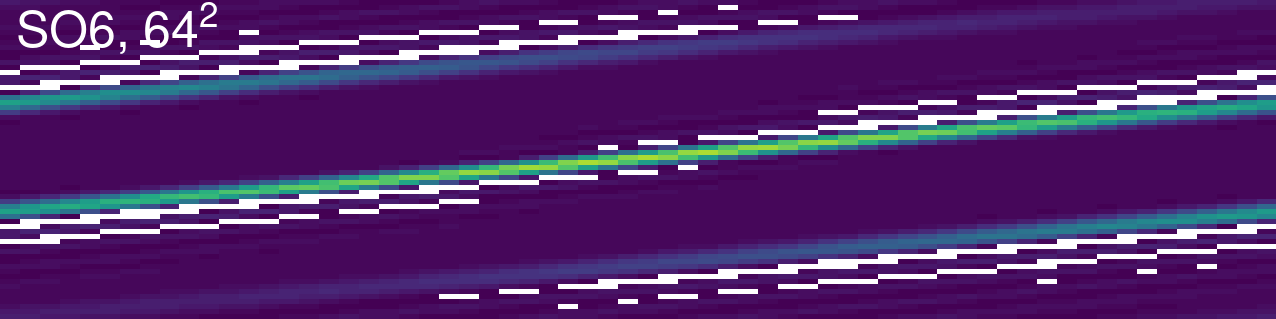}\\
    \rotatebox{90}{\hspace{0.037\linewidth}$y$} &
    \includegraphics[width=0.32\linewidth]{Figures/ao2_x64.png} &
    \includegraphics[width=0.32\linewidth]{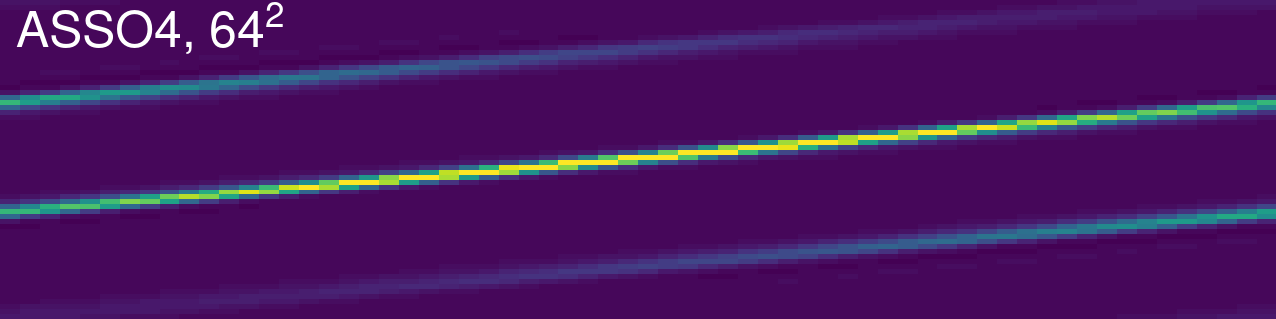} &
    \includegraphics[width=0.32\linewidth]{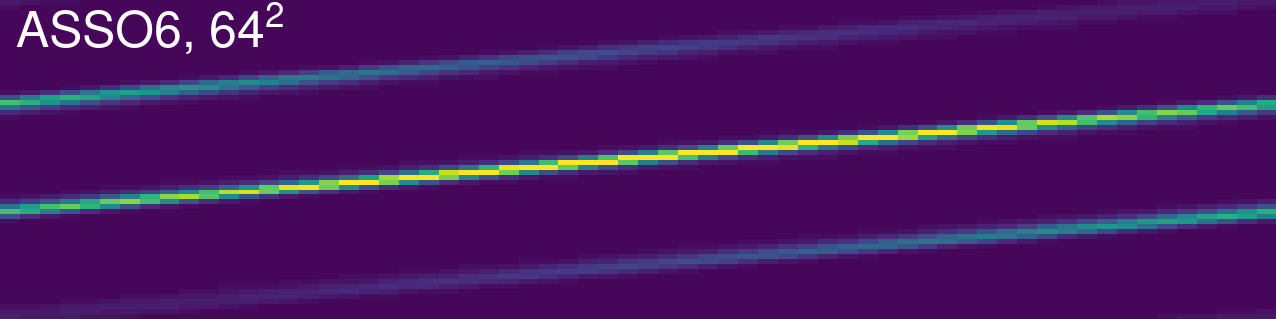} \\
    & $x$ & $x$ & $x$
    \end{tabular}
    \caption{Numerical solutions of Eq.~\ref{eq:ConvDiff} with $\kappa_\parallel=10$ at $t=10$ computed using conventional finite volume (top row, FV), anti-symmetry finite volume (second row, ASFV), conventional support operator (third row, SO), and anti-symmetry support operator (fourth row, ASSO) at $(n_x,n_y)=(64,64)$. For each method, simulations are carried out using 2nd, 4th, and 6th order accurate stencils (left, center, and right columns). Identical color scale is used for all images.}
    \label{fig:profiles2}
\end{sidewaysfigure}
We carried out simulations using the anti-symmetry finite volume (ASFV), anti-symmetry support operator (ASSO), conventional finite volume (FV), and conventional support operator (SO) methods. The grid size was varied between $(n_x,n_y)=(32,32)$ up to $(n_x,n_y)=(1024,1024)$, and 2nd, 4th, and 6th accuracy order stencils were used. This exercise allows us to confirm the correct verification of each numerical scheme by comparison against the Fourier series solution. Figure~\ref{fig:Convergence} shows the results of the resolution scans in log-log scale. The left column plots show the infinity norm error, while the right column show the square norm error. Each row corresponds to a different accuracy order, as marked on each figure. The infinity norm error corresponds to the difference between the maximum value of the computed solution and its true value. Asymptotic convergence is achieved for all the methods at all accuracy orders for grids larger than $(n_x,n_y)=(256,256)$, \emph{i.e.} provided that the grid can resolve the Fourier series solution. In general, the FV method performs the worst, and the ASSO method performs the best. The anti-symmetry FV algorithm shows significant improvement over its conventional FV counterpart. The two SO-based algorithms perform very well at all orders, and appear to reach their asymptotic convergence regime faster than the FV based methods.

Figure~\ref{fig:profiles1} is provided as a visual aid to support these observations. From left to right cases computed using different methods at grid sizes, $(n_x,n_y)=(64,64),(256,256)$, and $(1024,1024)$. The images are all rasterized using the simulated resolution, no interpolations are carried out. Identical color scale is used for all the figures, with white color indicating positivity violations ($f<0$). The five rows are, in order, the spectral solution, FV, ASFV, SO, and ASSO, as marked on the figure. A numerical suffix $(2,4,6)$ is added to note the accuracy order. The FV method simulations (second row) show significant perpendicular spread, which improves but does not disappear at higher resolutions. The ASFV simulations (third row) have less perpendicular spread. The SO simulations (fourth row) have little perpendicular spread, but we observe artifacts such as ringing and positivity violations with $f<0$. The ASSO simulations (fifth row) show even less perpendicular spread than the previous case, also accompanied by less ringing and no positivity issues.

Next, we inspect the solution quality for each algorithm at fixed resolution $(n_x,n_y)=(64,64)$ and different accuracy orders. The results are shown in Fig.~\ref{fig:profiles2}. The rows correspond to each algorithm (FV, ASFV, SO, ASSO), while the columns correspond to 2nd (left), 4th (center), and 6th (right) accuracy order simulations. The FV method (top row) improves marginally with higher order stencils. While the perpendicular spreading improves, we observe additional artifacts such as ringing and positivity violations, \emph{e.g.} for the FV4 and FV6 cases. The ASFV algorithm (second row) shows a marked improvement between 2nd and 4th order. Shown on the following row, the SO methods show little perpendicular spread. However, positivity violations and ringing still appear at high order. The last row (ASSO algorithm) does not reveal much improvement with increasing accuracy order, \emph{e.g.} the algorithm produces visually satisfactory solutions already at low resolution and low precision.

\begin{figure}[ht!]
    \centering
    \includegraphics[height=0.25\linewidth]{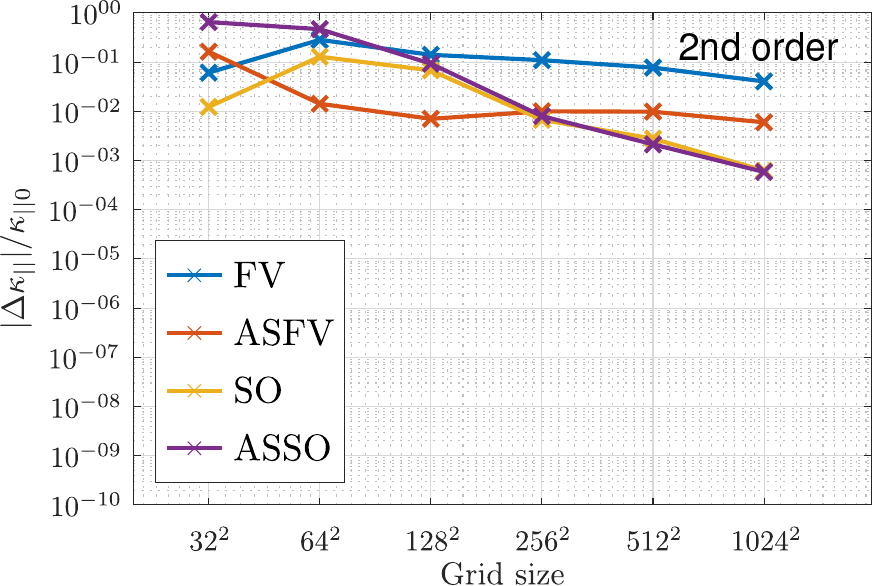}
    \includegraphics[height=0.25\linewidth]{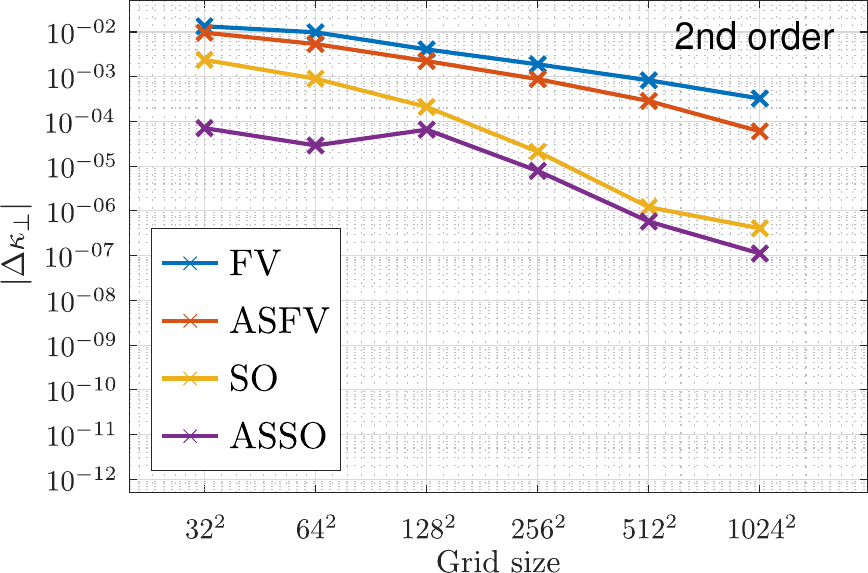}\\
    \includegraphics[height=0.25\linewidth]{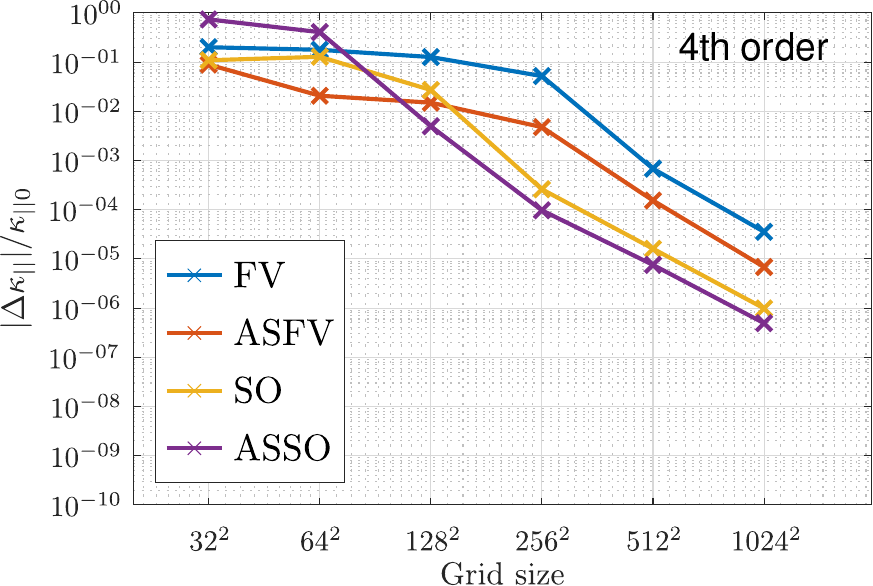}
    \includegraphics[height=0.25\linewidth]{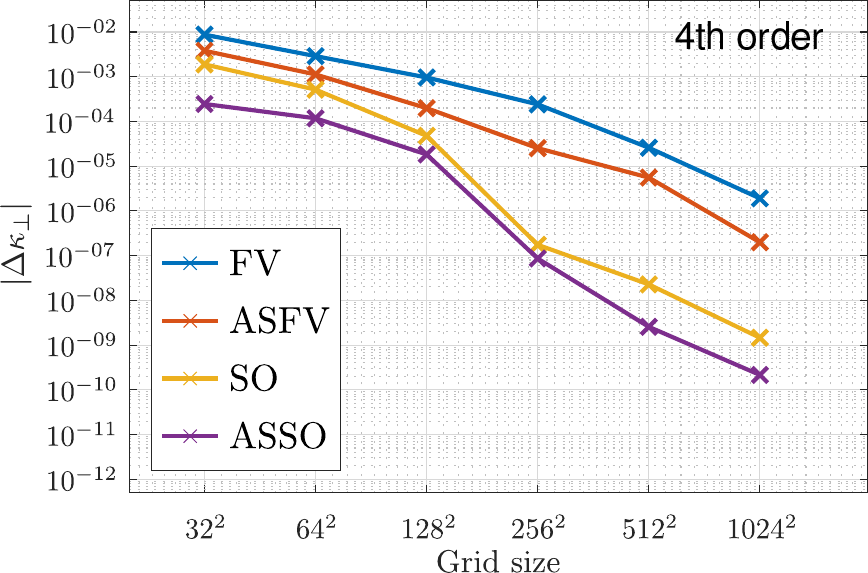}\\
    \includegraphics[height=0.25\linewidth]{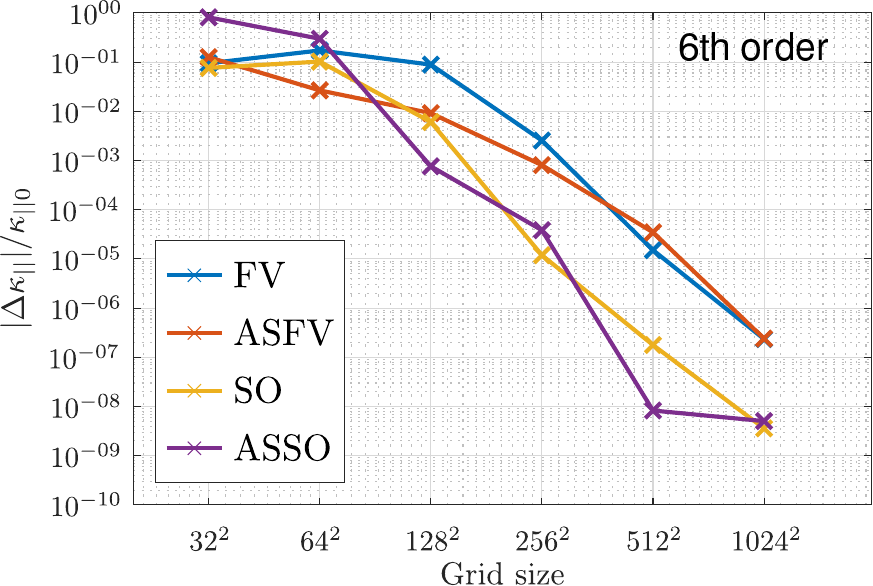}
    \includegraphics[height=0.25\linewidth]{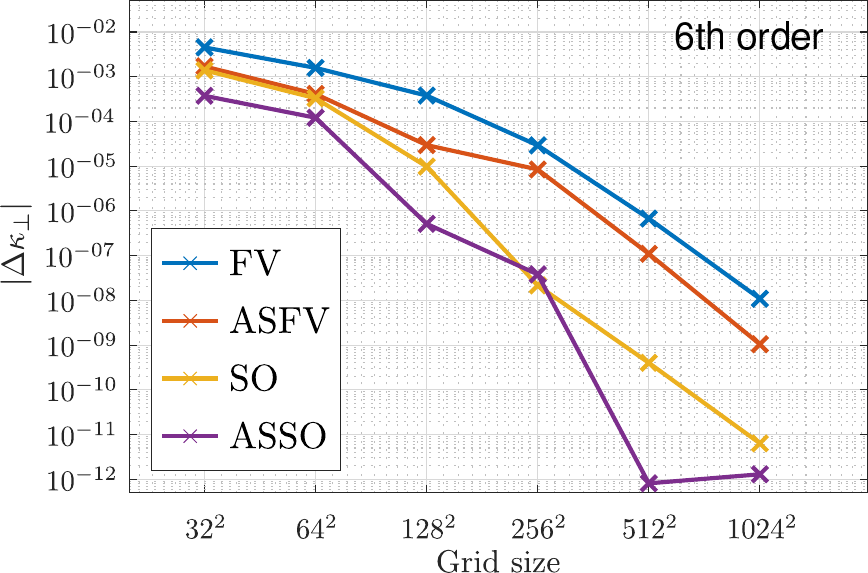}
    \caption{Computed relative error on parallel diffusion (left column) and absolute error on perpendicular diffusion (right column) in 2D simulations of parallel diffusion of a tilted Gaussian function. Anti-symmetry (AS) and conventional representations are used, implemented using finite volume (FV) and support operator (SO) methods of 2nd, 4th, and 6th order accuracy. Results are shown at $t=10$ with $\kappa_\parallel=10$, $q=3$.}
    \label{fig:EffectiveDiffusion}
\end{figure}

It is possible to provide a quantitative metric of the parallel and perpendicular spread for all the simulations. This is achieved by fitting a tilted 2D Gaussian on the simulation results, and then estimating the effective conductivities (Eqs.~\ref{eq:SpreadPar} and~\ref{eq:SpreadPerp}) from the fitted 
$\sigma_\parallel$ and $\sigma_\perp$. These estimates result from nonlinear fits of 2D data, which have residuals $\sim 10^{-10}$. The standard error for the Gaussian widths appears to be negligible. Using the covariance matrix for the nonlinear fit computation, their standard error ranges from $10^{-3}$ for the low resolution simulations to $\sim 10^{-10}$ for the high resolution simulations. The results are shown in Fig.~\ref{fig:EffectiveDiffusion}.  The left column shows the relative error in the parallel conductivity. All of the methods perform reasonably well and attained errors of a few percent or less at low resolutions. One interesting feature is that sometimes the ASSO method did not produce enough parallel diffusion at low grid sizes. Another observation is that the 2nd order FV method does not converge particularly well even for this smooth and low wave number case.

\begin{sidewaysfigure}[!]
    \centering
    \begin{tabular}{c@{\hskip2mm}c@{\hskip0.5mm}c@{\hskip0.5mm}c}
    \rotatebox{90}{\hspace{0.037\linewidth}$y$} &
    \includegraphics[width=0.32\linewidth]{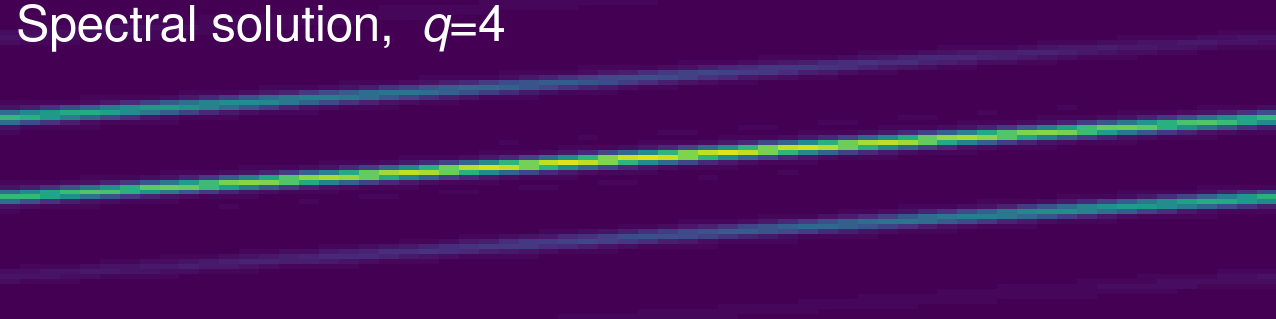} &
    \includegraphics[width=0.32\linewidth]{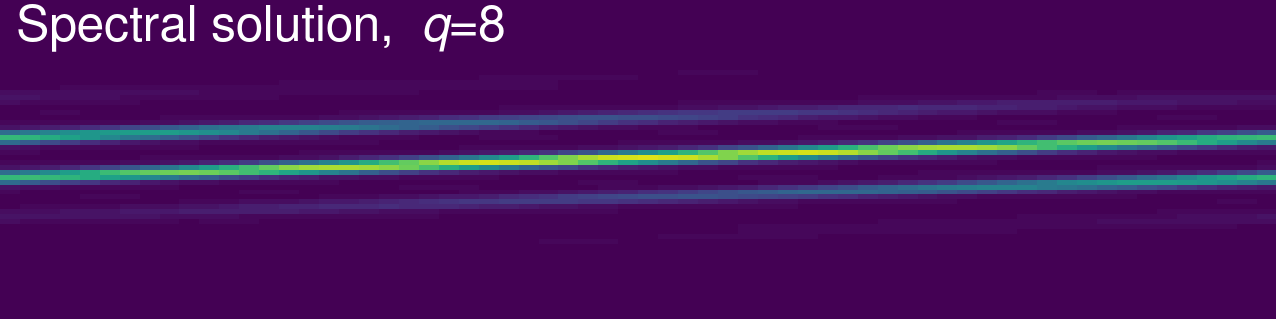} &
    \includegraphics[width=0.32\linewidth]{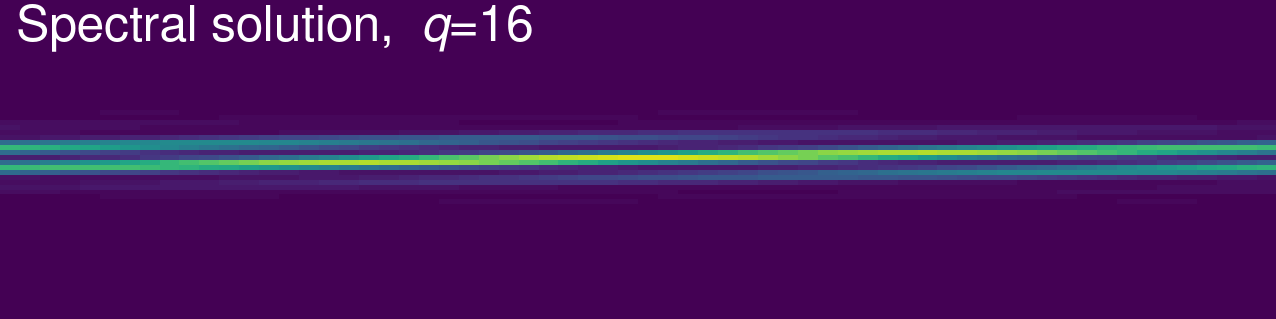}\\
    \rotatebox{90}{\hspace{0.037\linewidth}$y$} &
    \includegraphics[width=0.32\linewidth]{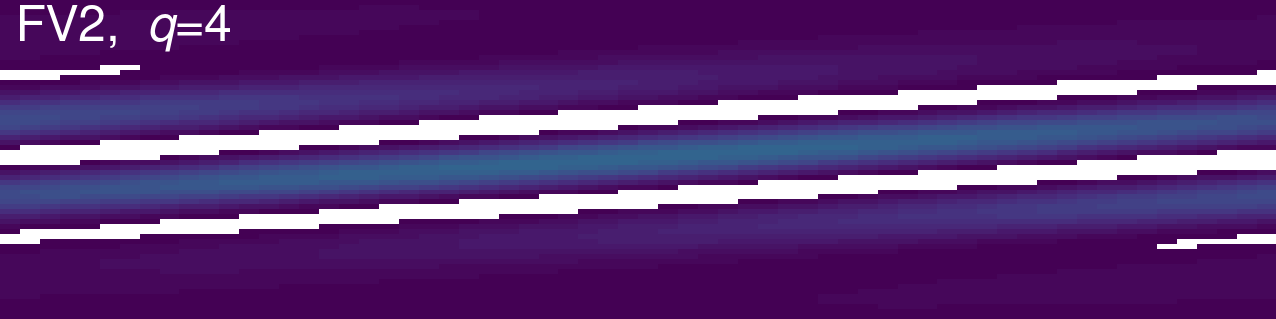} &
    \includegraphics[width=0.32\linewidth]{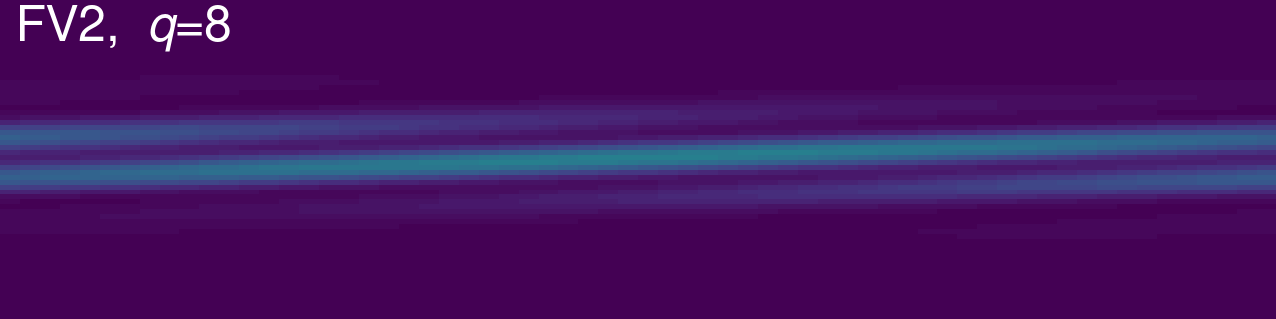} &
    \includegraphics[width=0.32\linewidth]{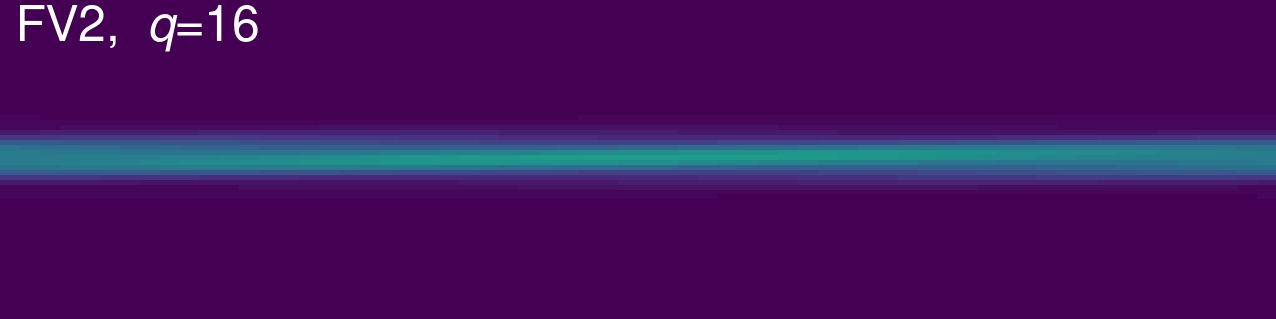}\\    
    \rotatebox{90}{\hspace{0.037\linewidth}$y$} &
    \includegraphics[width=0.32\linewidth]{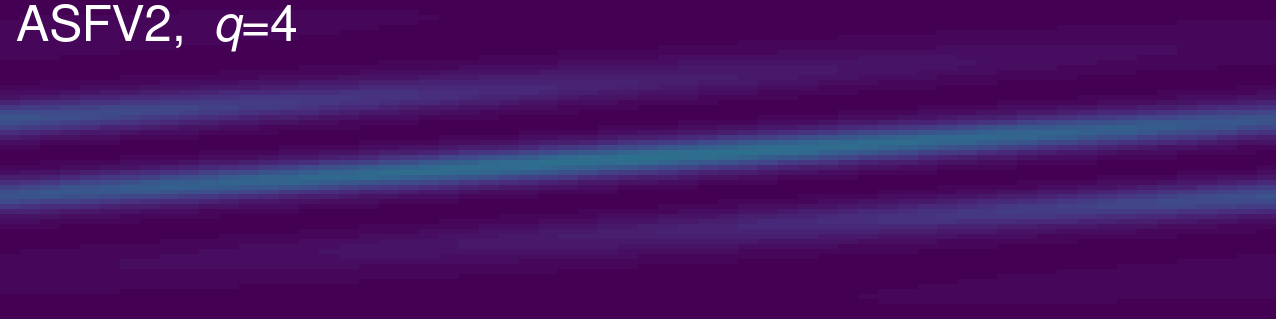} &
    \includegraphics[width=0.32\linewidth]{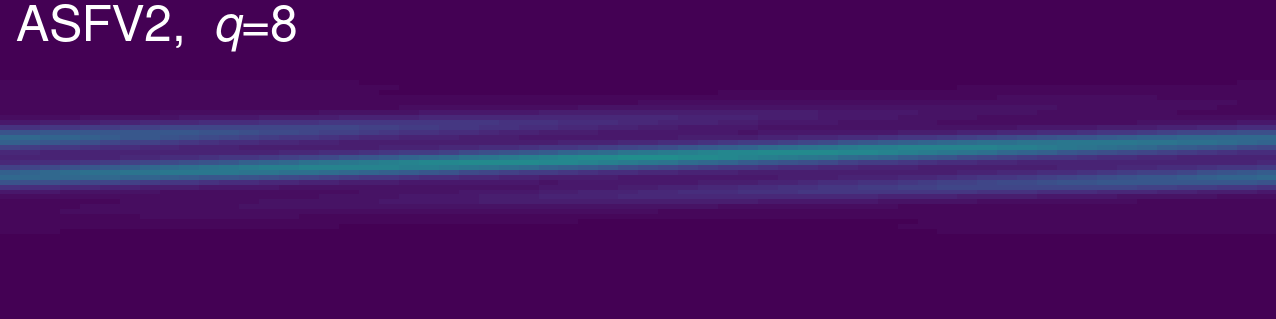} &
    \includegraphics[width=0.32\linewidth]{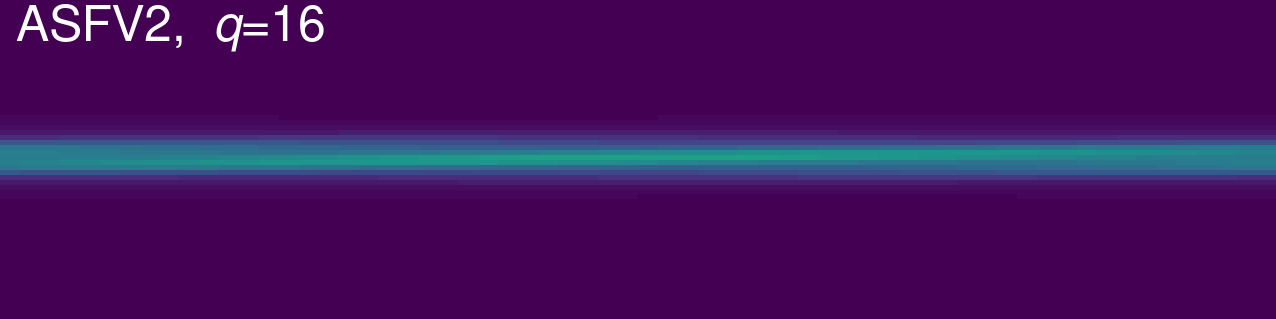}\\
    \rotatebox{90}{\hspace{0.037\linewidth}$y$} &
    \includegraphics[width=0.32\linewidth]{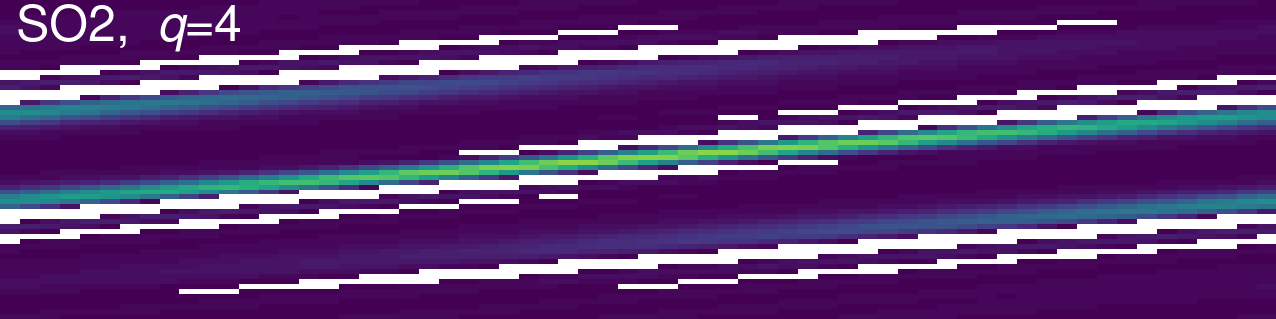} &
    \includegraphics[width=0.32\linewidth]{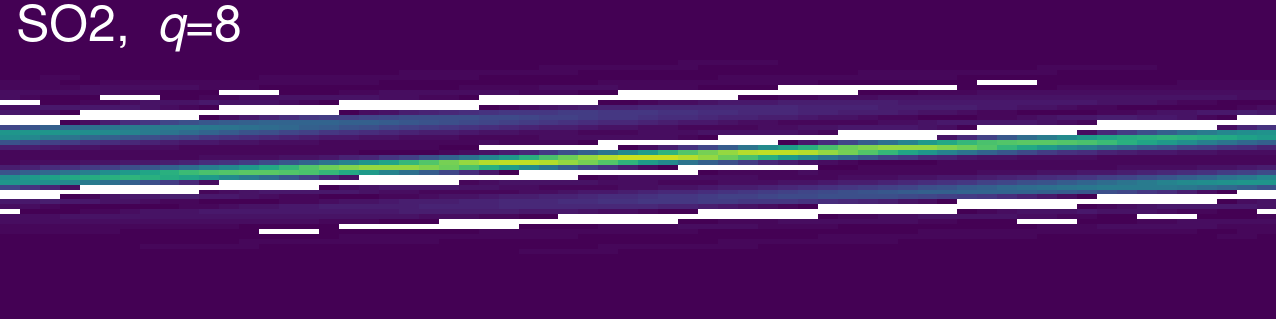} &
    \includegraphics[width=0.32\linewidth]{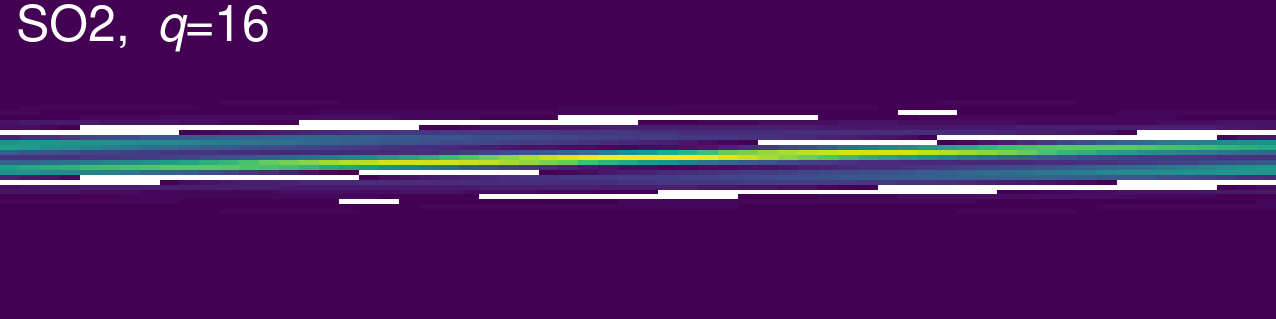}\\
    \rotatebox{90}{\hspace{0.037\linewidth}$y$} &
    \includegraphics[width=0.32\linewidth]{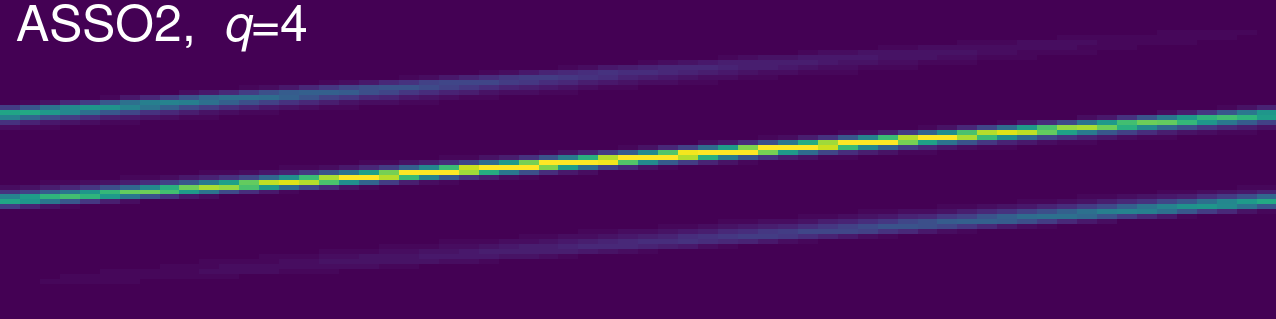} &
    \includegraphics[width=0.32\linewidth]{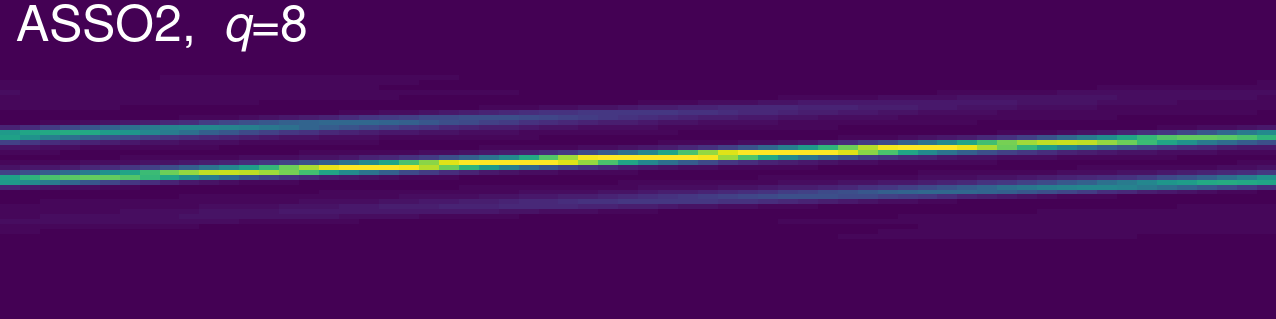} &
    \includegraphics[width=0.32\linewidth]{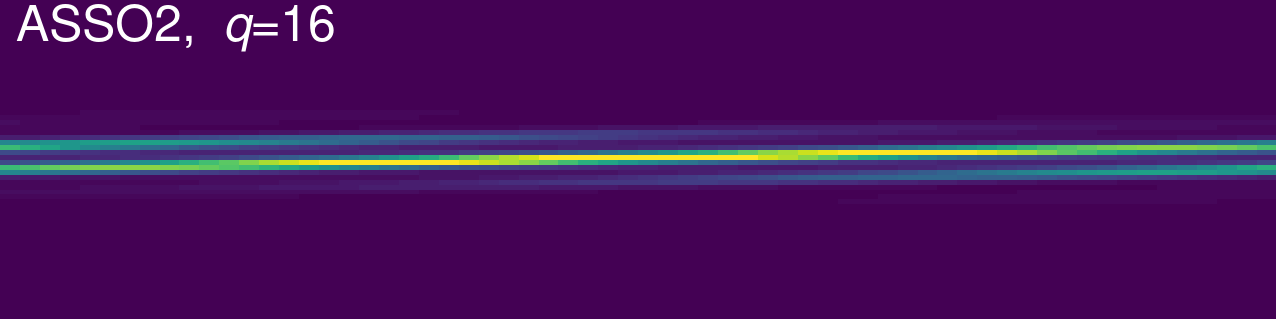} \\
    & $x$ & $x$ & $x$
    \end{tabular}
    \caption{Numerical solution of Eq.~\ref{eq:ConvDiff} with $\kappa_\parallel=10$ at $t=10$ in a planar constant magnetic field with $q=4,8,16$ (left, center, right columns). Solutions are computed using Fourier series (Eq.~\ref{eq:Solution2D}) (top row),
    conventional finite volume (second row, FV),
    anti-symmetry finite volume (third row, ASFV),
    conventional support operator (fourth row, SO),
    and
    anti-symmetry support operator (fifth row, ASSO) algorithms, all using 2nd order accurate stencils. Grid resolution is $(n_x,n_y)=(64,64)$. Identical color scale is used for all images.}
    \label{fig:profiles3}
\end{sidewaysfigure}

The right column of Fig.~\ref{fig:EffectiveDiffusion} shows the absolute error of the perpendicular spread over the elapsed time is hastily interpreted as an effective perpendicular diffusion coefficient. The SO-based algorithms perform best, but ASSO is almost always better, especially at low order and low resolution.  The value $\sigma_\perp \sim 10^{-5}$ is comparable to the artificial viscosity used by turbulence codes in the perpendicular direction. All in all, the ASSO algorithm reduces the perpendicular error by a factor 2--10, w.r.t. SO simulations, and by a factor $10^2$--$10^3$ w.r.t. conventional FV method. 

Next, we investigate the minimum magnetic field tilt allowed by each method, \emph{i.e.} the maximum $q$ value allowed for a given grid size. This exercise addresses the implementation of magnetic nulls using non-field aligned geometry or algorithms. Simulations are carried out with resolution $(n_x,n_y)=(64,64)$, with $q=4,8,16$, 2nd order accurate stencils. This resolution is chosen to The results are shown in Fig.~\ref{fig:profiles3} together with a reference spectral solution. Solution quality roughly corresponds to the expectations set by the previous accuracy and profile results. While all methods are able to resolve the $q=4$ case, the FV method, in particular, already shows significant degradation at $q=8$ (8 points per wavelength) and breaks down completely for $q=16$. The ASFV results are somewhat improved with respect to FV. The SO results do not show eddy merging, but they develop unphysical sidebands, which are again accompanied by positivity violation issues as observed in the $q=3$ case above. It is speculated that the sidebands emerge due to excessive action of the low-pass filter terms at high $k$. The results of the ASSO algorithm resemble the spectral solutions quite closely. 


\subsection{Parallel diffusion in a non-uniform 2D magnetic field}
The purpose of the present section is to verify whether the results from the previous subsection carry, at least qualitatively, to a 2D case where the magnetic field is not uniform. We use Cartesian geometry, $L_x=L_y=2\pi$, $q=3$, with insulating boundary conditions. The magnetic field and the initial condition are given by
\begin{align}
    B_x &= -\rho\frac{\cos{\theta}}{q},~~B_y = \rho\frac{\sin{\theta}}{q},\label{eq:Bfield2D}\\
    f  &= 10\exp{ \left\{-\frac{\left(\rho-\rho_0\right)^2}{2\sigma_{\perp 0}^2} -\frac{\left(\theta - \theta_0\right)^2}{2\sigma_{\parallel 0}^2}\right\} } + f_{\mathrm{bg},0}.
\end{align}Here $\rho = \sqrt{x^2+y^2}$ is the radial coordinate and $\theta=\tan^{-1}{(y/x)}$ is the poloidal angle. The Gaussian is initialized with initial location $(\rho_0,\theta_0)=(2,1)$, initial widths $\sigma_{\perp 0} = 0.1$ and $\sigma_{\parallel 0}=0.2$, and a small background constant $f_{\mathrm{bg},0}=5\times10^{-3}$. The grid size is varied from $(n_x,n_y)=(32,32)$ to $(n_x,n_y)=(512,512)$, and 2nd, 4th, and 6th accuracy order stencils are used for each scheme. Time integration is carried out with an RK4 method, with $\Delta=10^{-3}$ at the lowest resolution and adjusted with the CFL condition for larger grids. The simulations are evolved to $t=10$.

\begin{figure}
    \centering
    \begin{tabular}{c@{\hskip2mm}c@{\hskip0.5mm}c}
        \rotatebox{90}{\hspace{0.14\linewidth}$y$} &
        \includegraphics[width=0.3\linewidth]{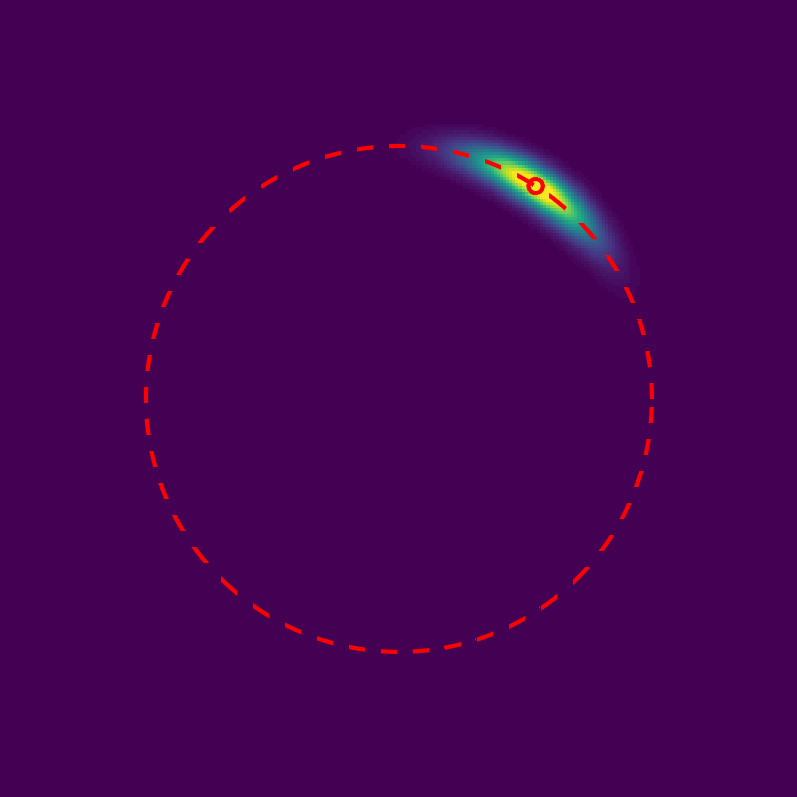} &  
        \includegraphics[width=0.3\linewidth]{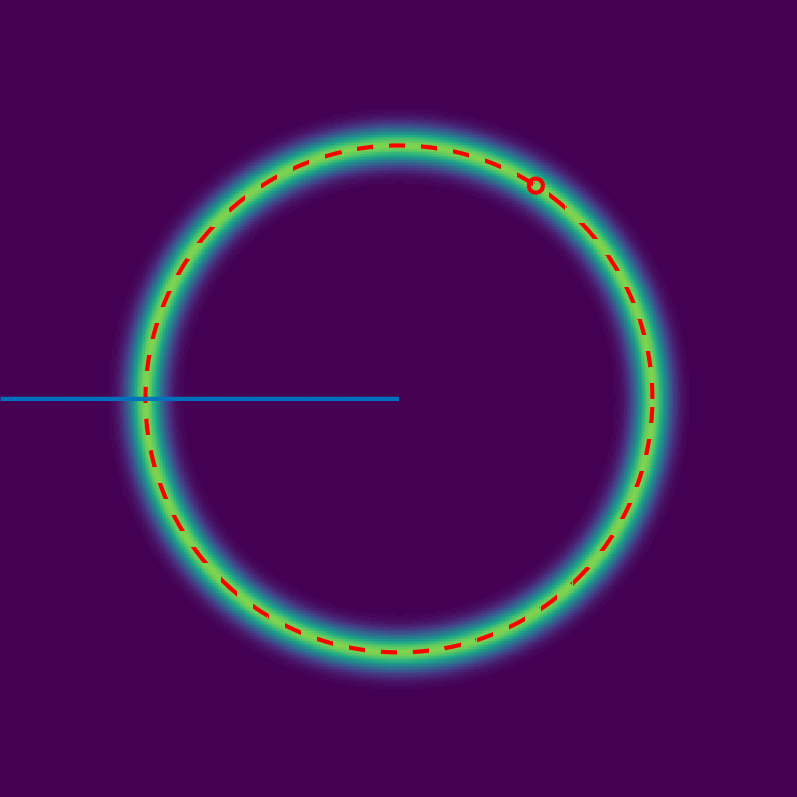}
        \\
        & $x$ & $x$ 
    \end{tabular}
    \caption{Simulation of parallel diffusion of a narrow Gaussian pulse in a circular magnetic field (Eq.~\ref{eq:Bfield2D}). After $t=10$, the pulse spreads into an annulus. Left panel: Initial state. Right panel: Final state at $t=10$, with a blue line indicating the angle where profile fitting is carried out for Eq.~\ref{eq:RadialFit}. Simulation carried out using 6th order accuracy anti-symmetry support operator (ASSO), $(n_x,n_y)=(512,512)$.}
    \label{fig:Ring2D}
\end{figure}

\begin{figure}[!t]
    \centering
    \includegraphics[width=0.32\linewidth]{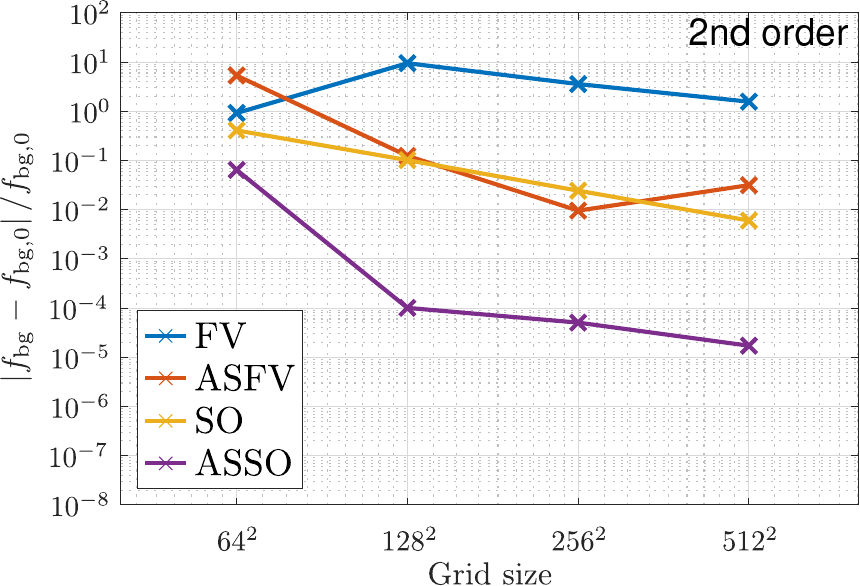}
    \includegraphics[width=0.32\linewidth]{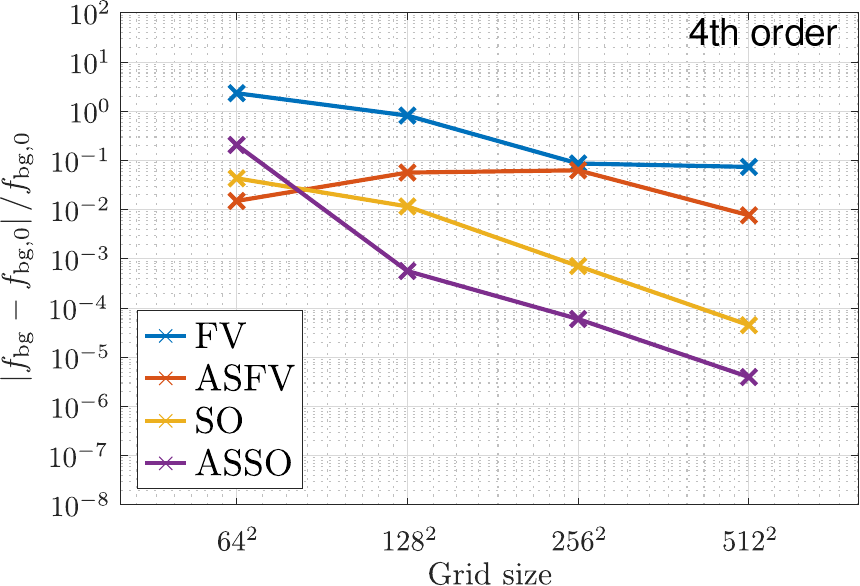}
    \includegraphics[width=0.32\linewidth]{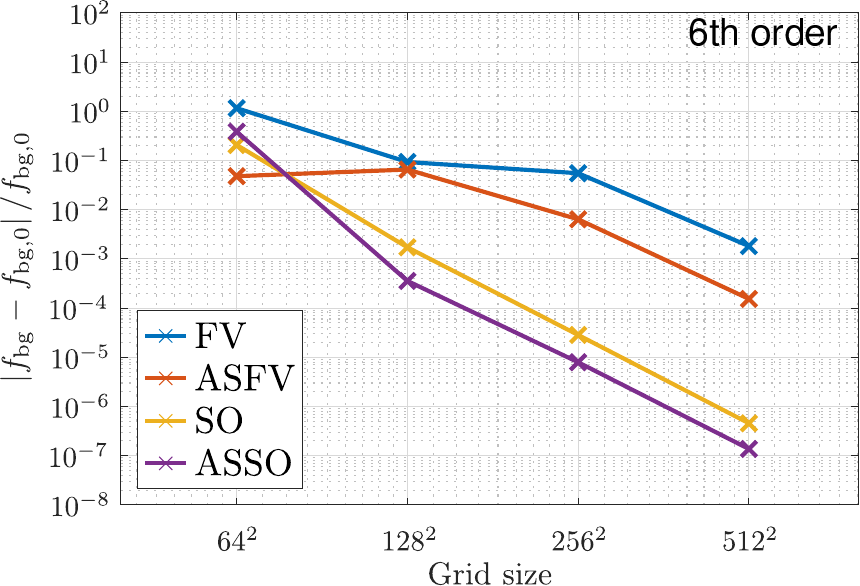}\\
    \includegraphics[width=0.32\linewidth]{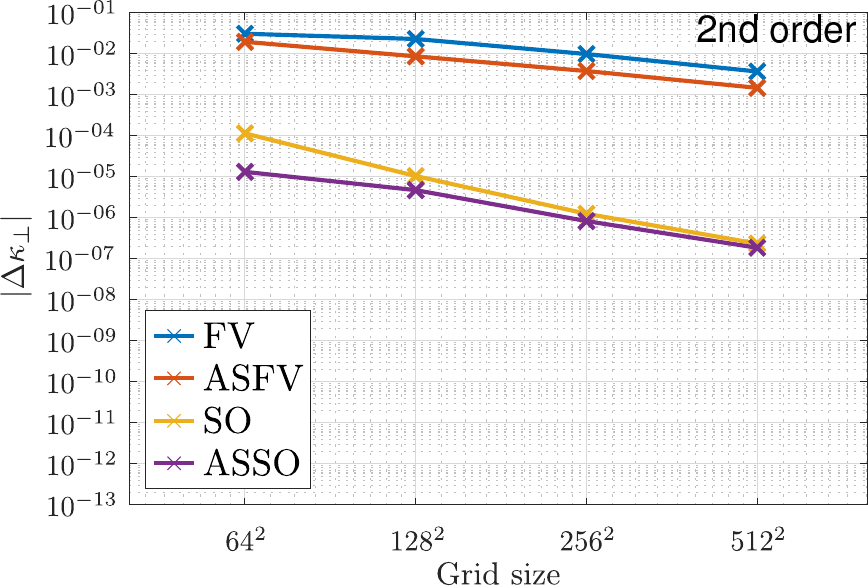}
    \includegraphics[width=0.32\linewidth]{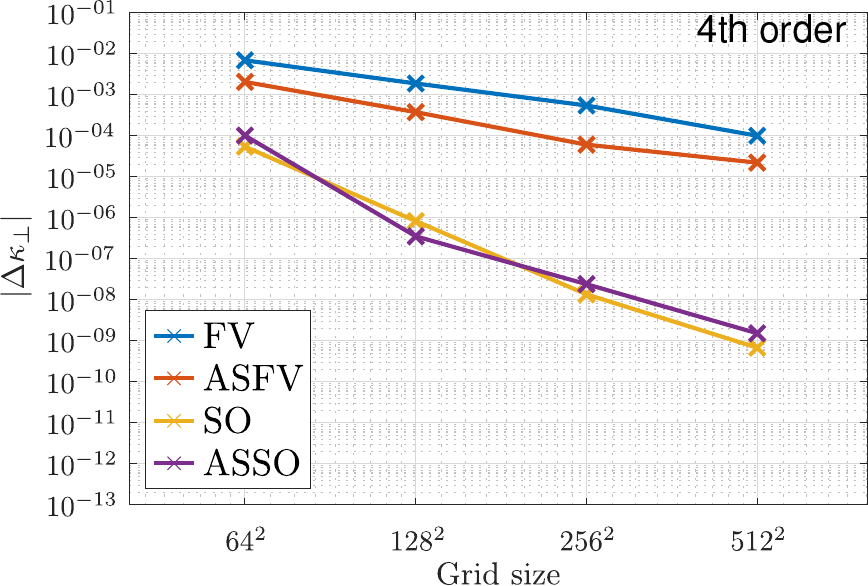}
    \includegraphics[width=0.32\linewidth]{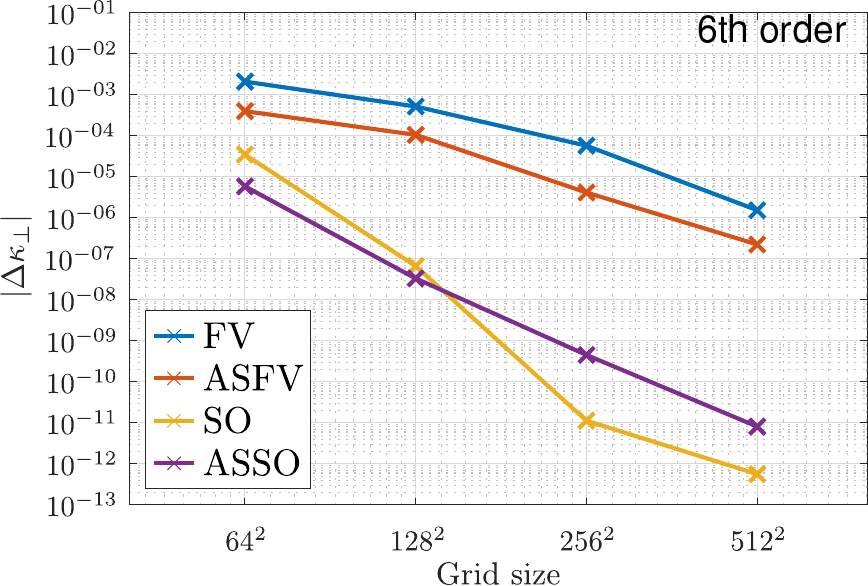}
    \caption{Profile fit results for simulations of a Gaussian diffusing in a circular  magnetic field. Top row: relative error in background value, $f_{\mathrm{bg}}$. Bottom row: perpendicular diffusion $\kappa_\perp$, estimated from the final width of the Gaussian pulse. Simulation carried out using 2nd, 4th, and 6th order accuracy methods (left, center, right).}
    \label{fig:Ring2DError}
\end{figure}

The expected final profile (Fig~\ref{fig:Ring2D}, right panel) is a poloidally uniform ring around $\rho_0=2$ (red dashed line) with a radial Gaussian envelope of width $\sigma_\perp=\sigma_{\perp 0}=0.1$. The numerical results are verified using nonlinear fitting. We sometimes observe anomalies of numerical origin, depending on the representation and discretization stategy used. Simulations based on the conventional formulation (FV and SO methods) sometimes yield profiles with considerable positivity violations or ringing. The algorithms based on the FV discretization (FV and ASFV) produce wider Gaussian profiles, which result from artificial perpendicular diffusion. In combination with the insulating boundary conditions, this effect leads to material accumulation at the domain edges. Additionally, accumulation around the center of domain is occasionally found. These two effects, in particular, motivated the use of additional parameters to improve the fit around the Gaussian peak. Assuming negligible poloidal dependence, the following fitting model is used:
\begin{align}
    f(\rho) & = A_0 \exp{\left\{-\frac{(\rho-\rho_0)^2}{2\sigma_\perp^2}\right\}}
    + A_1 J_0\left(c_0 \frac{ \rho}{\rho_{\mathrm{max}}}\right) + A_2 J_2\left(c_2 \frac{ \rho}{\rho_{\mathrm{max}}}\right) + f_{\mathrm{bg}}. \label{eq:RadialFit}
\end{align}The value $c_0=11.791$ is the 4th zero of $J_0$, while $c_2=3.0542$ is the 1st maximum of $J_2$, the zeroth and second Bessel functions of the first kind. The function $J_0$ is employed to capture material accumulation near the center, while $J_2$ is employed to capture accumulation near the edge. Eq.~\ref{eq:RadialFit} results from the intuition that, as a consequence of numerical error, radially symmetric beating could occur superimposed with a radial Gaussian solution. In a well resolved numerical solution, we expect $\sigma_\perp\approx 0.1$, $\rho_0\approx2$, $A_1\approx 0$, $A_2\approx 0$, $f_{\mathrm{bg}}\approx5\times10^{-3}$. We do not attempt to capture ringing around the Gaussian. A single radial profile is fit at $\theta=\pi$ using data at $y=0$ defined in the $(x,y)$ grid -- marked as a blue line on the right panel of Fig.~\ref{fig:Ring2D}.

Fig~\ref{fig:Ring2DError} illustrates our findings on two key results for which the solution is known -- the fitted value of $f_{\mathrm{bg}}$ (top row) and an estimate of the effective perpendicular conductivity, $k_\perp$, obtained by using the fitted perpendicular width $\sigma_\perp$ in Eq.~\ref{eq:SpreadPerp} (bottom row). The results are shown for 2nd, 4th, and 6th order methods (left, center, and right panels). The first result ($f_{\mathrm{bg}}$) is a measure of how much material is overall leaking into the background, while the second one indicates how well the method works on the initial Gaussian pulse itself. The trends found here are comparable to those shown in the previous subsection for a filament diffusing along a constant magnetic field. The FV method struggles to produce usable results even at higher order. At low to medium resolution, the profiles computed using this method are dominated by artificial diffusion, which is of order $\mathcal{O}(1)$. Although the ASFV method results in significantly better background values, it fails to reduce the artificial diffusion to acceptable levels. The SO-based methods both perform very well, with SO sometimes outperforming ASSO at high order. 

\begin{figure}
    \centering
    \begin{tabular}{c@{\hskip2mm}c@{\hskip0.5mm}c}
    \rotatebox{90}{\hspace{0.14\linewidth}$y$} &
    \includegraphics[width=0.3\linewidth]{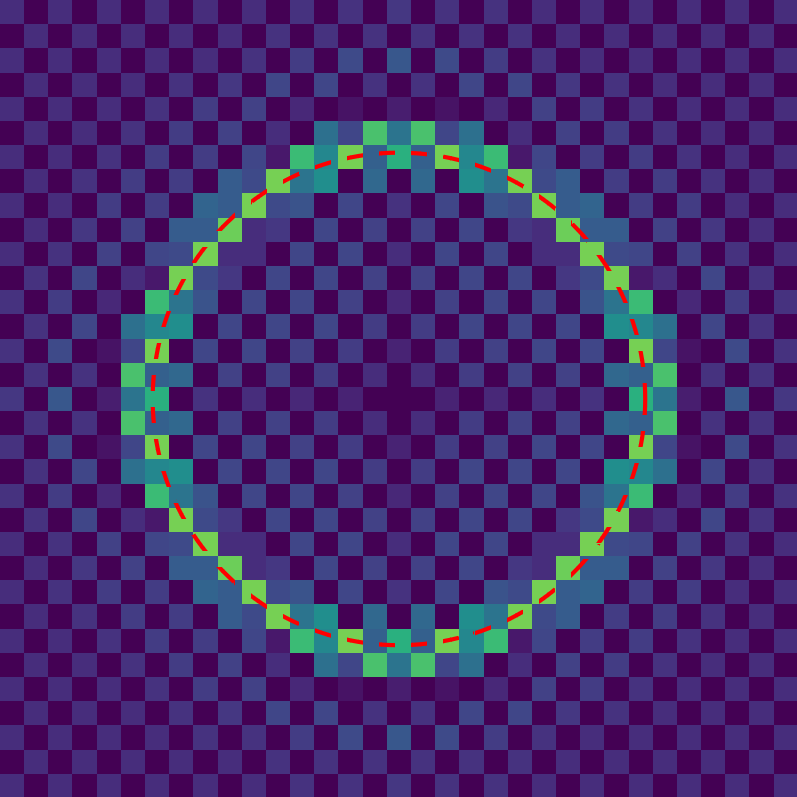} &
    \includegraphics[width=0.3\linewidth]{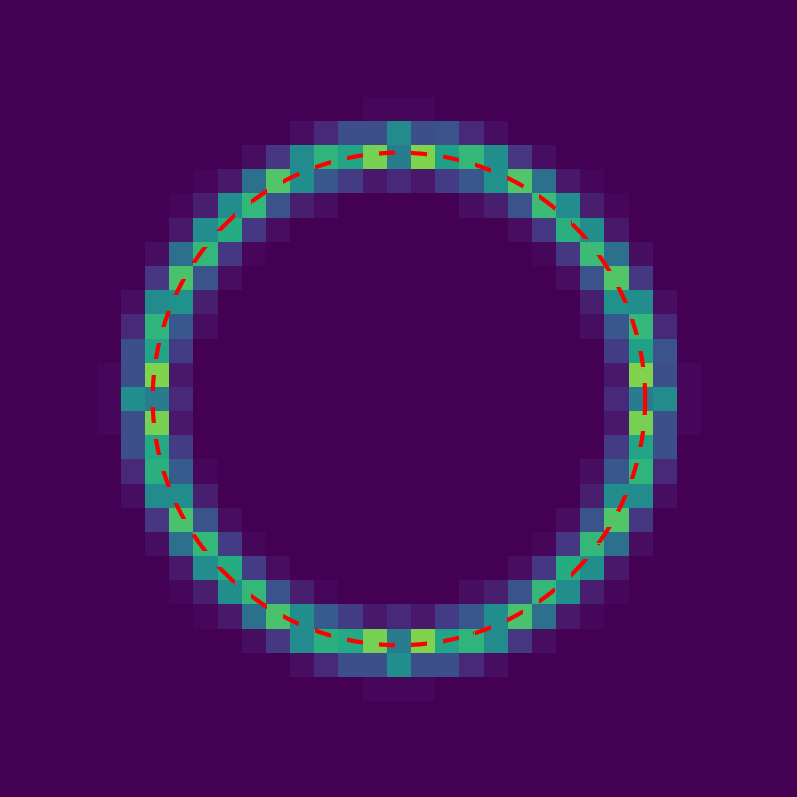}\\
    & $x$ & $x$
    \end{tabular}
    \caption{Final state ($t=10$) for simulation of a narrow Gaussian pulse diffusing around a circular field, spatial resolution $(n_x,n_y)=(32,32)$. Solutions are computed using  2nd order support operator method (SO, left) or anti-symmetry support operator method (ASSO, right).}
    \label{fig:Ring2Dx32}
\end{figure}

Even with the smaller error occasionally found using SO vs.~ASSO, the latter method still brings some interesting advantages. The profiles computed with this method had negligible ringing, and there are no positivity violations. The most significant shortcoming of the SO method was that it sometimes results in significant checkerboarding, in particular at low resolution $(n_x,n_y)=(32,32)$ (Fig.~\ref{fig:Ring2Dx32}). At this resolution, the Gaussian radial profile is significantly under resolved, with just 3 or 4 points for the peak of the waveform. We speculate that the SO checkerboarding effect originates from excessive filtering at high wave number -- the method does not damp the highest $k_x$'s and $k_y$'s in the system due to a large diffusion shortfall (recalling Eq.~\ref{eq:Response2DSO} and its corresponding panel on Fig.~\ref{fig:Response2D}). Checkerboarding occurs with the SO method regardless of the accuracy order, but the anti-symmetry SO method does not appear to suffer from this issue. The FV based methods do not manifest this issue either, as the artificial spreading of the Gaussian dominates the dynamics. 

\subsection{Parallel diffusion in a screw-pinch}
We now verify the numerical properties of the algorithms in a simple 3D geometry with a varying magnetic field. The test case originates from a paper discussing the parallel diffusion accuracy of the FCI method~\cite{Stegmeir2023}. The numerical settings are matched as closely as possible. We use Cartesian geometry, $L_x=L_y=0.5$, $L_z=2\pi$, $q=3$. The time step is fixed at $\Delta t=2.5\times10^{-4}$, and the planar resolution is $(n_x,n_y)=(125,125)$, which results in $\Delta x=\Delta y=4\times10^{-3}$. The parallel conductivity is $\kappa_\parallel=1$, while $\kappa_\perp=0$. Spatial derivatives are computed with the conventional and anti-symmetry representations, with the support operator method as a discretization strategy (SO and ASSO, respectively). Based on the results on the previous section, the FV methods are not attempted, since they produce significant error. Both 2nd and 4th order SO and ASSO methods are used, to allow a close comparison with \cite{Stegmeir2023} -- the conventional SO discretization is equivalent to the GU2 and GU4 schemes in \cite{Stegmeir2023}. The magnetic field and the initial condition are given by
\begin{align}
    B_x &= -\rho\frac{\cos{\theta}}{q},~~B_y = \rho\frac{\sin{\theta}}{q},~~B_z=1,\\
    f  &= \exp{ \left\{-\frac{\left(\rho-\rho_0\right)^2}{2\sigma_{\perp 0}^2} -\frac{\left(l_\parallel-l_{\parallel 0} \right)^2}{2\sigma_{\parallel 0}^2}\right\} } + f_{\mathrm{bg},0}.
\end{align}Here $\rho = \sqrt{x^2+y^2}$ is the radial coordinate, $l_\parallel$ describes the distance along the helicoidal magnetic field line, and $f_{\mathrm{bg},0}=5\times10^{-3}$. The Gaussian has initial widths $\sigma_{\perp 0} = 0.025$ and $\sigma_{\parallel 0}=\pi$. We pick $(x,y,z)=(0.15,0,0)$ as the start of the field line ($\rho_0=0.15, \theta_0=0,z=0$), and trace the field line location using a parameterized helical function. Although the perpendicular distance can be obtained from the gradient of the helical function, the perpendicular part of the initial condition is simplified such that it lays on the $x$--$y$ plane. This is a common simplification used in \cite{Stegmeir2018,Zhu2018,Giacomin2022,Dudson2024} and other turbulence codes, justified by the small tilt of the perpendicular plane. 

\begin{figure}[!t]
    \centering
    \begin{tabular}{c@{\hskip2mm}c@{\hskip0.5mm}c@{\hskip0.5mm}c}
    \rotatebox{90}{\hspace{0.12\linewidth}$y$} &
    \includegraphics[width=0.26\linewidth]{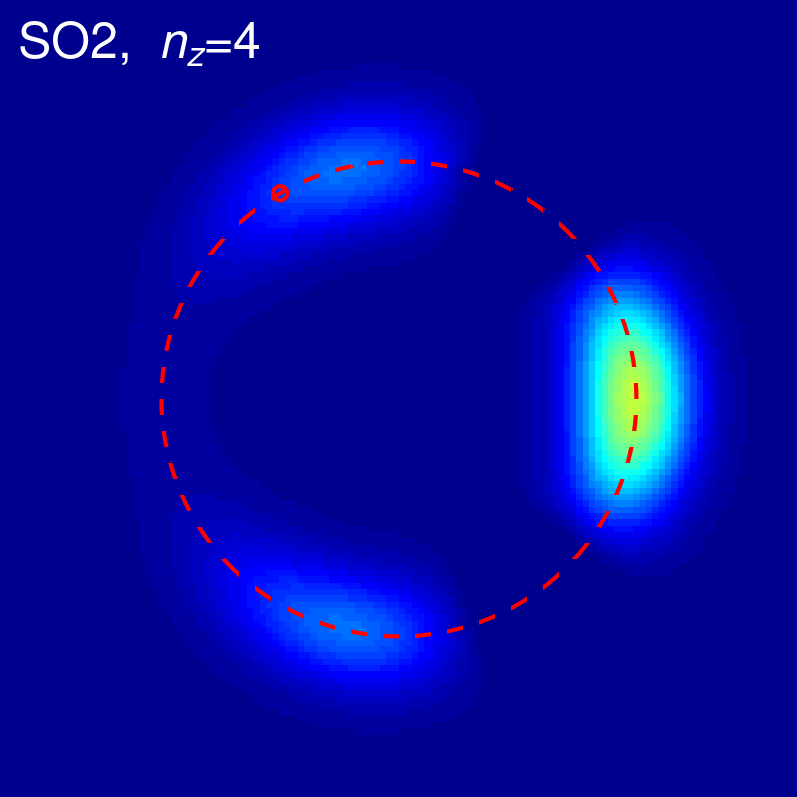} &
    \includegraphics[width=0.26\linewidth]{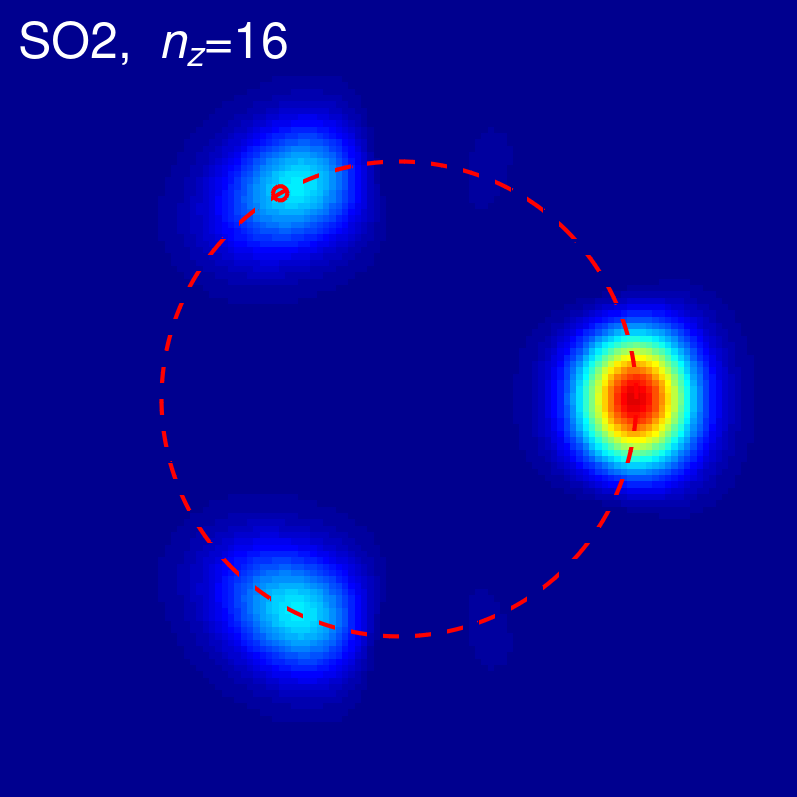} &
    \includegraphics[width=0.26\linewidth]{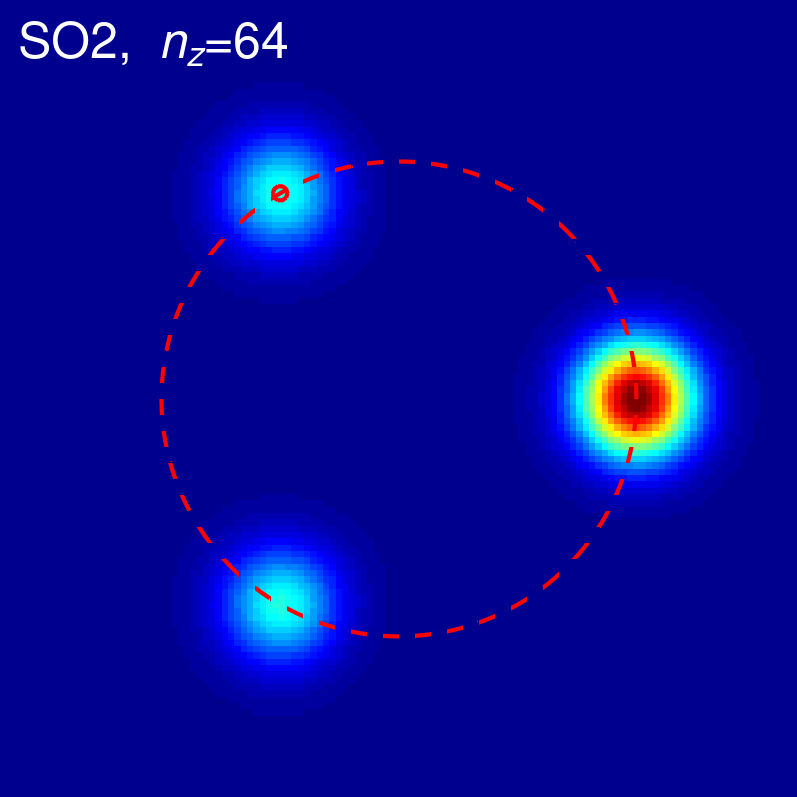}\\
    \rotatebox{90}{\hspace{0.12\linewidth}$y$} &
    \includegraphics[width=0.26\linewidth]{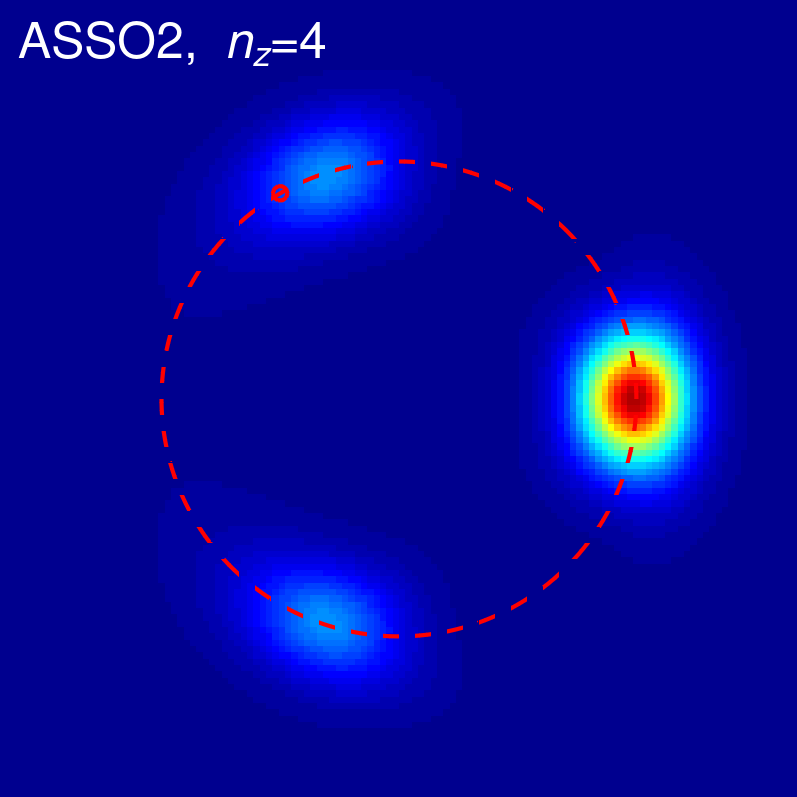}&
    \includegraphics[width=0.26\linewidth]{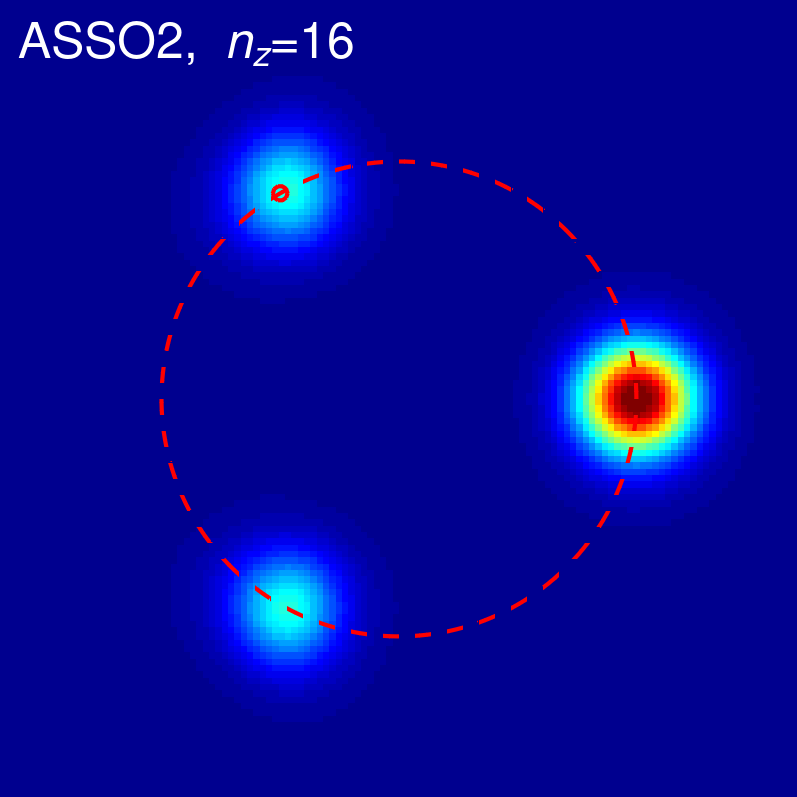}&
    \includegraphics[width=0.26\linewidth]{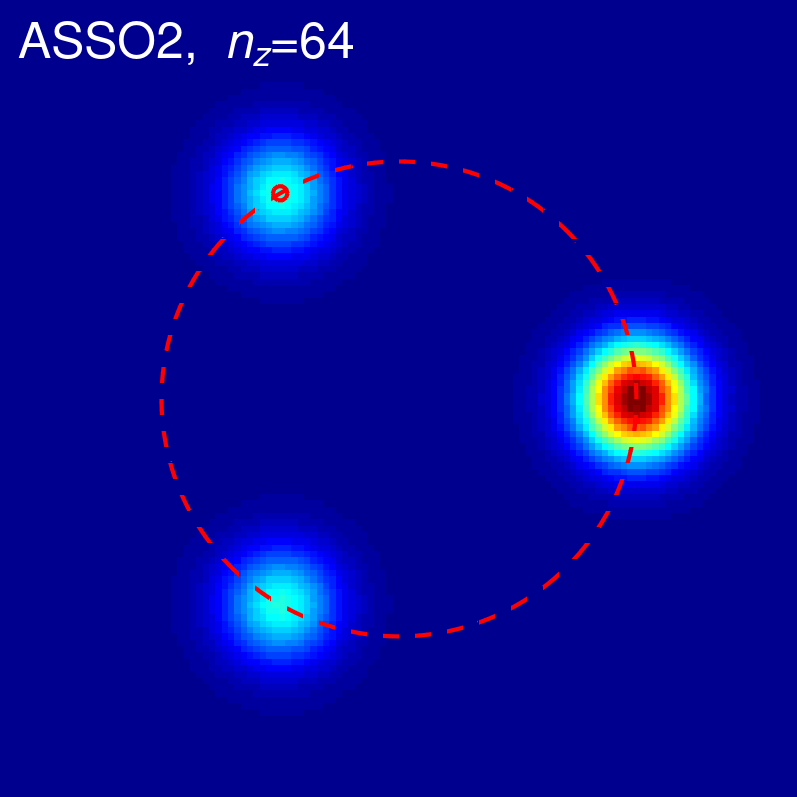}\\
    & $x$ & $x$ & $x$
    \end{tabular}
    \caption{Snapshots of 3D parallel diffusion for a field aligned blob at $z=0$, $t=5$, computed using 2nd order support operator (SO2, top row) and anti-symmetry support operator (ASSO2, bottom row) methods. Axial resolutions of 4, 16, 64 points shown on the left, center, right. Time step is $\Delta t=2.5\times10^{-4}$, $(n_x,n_y)=(125,125)$, $q=3$. The red line indicates $\rho=0.15$.}
    \label{fig:ScrewPinch}
\end{figure}

First, a minimal verification of the time stepping limits is performed. Using $(n_x,n_y,n_z)=(125,125,32)$, the maximum timesteps for each scheme are $\Delta t\approx1.75\times10^{-3}$ (SO2) and $\Delta t\approx1.7\times10^{-3}$ (ASSO2) for RK4. The steps are a factor of two longer steps than reported in Ref.~\cite{Stegmeir2023}, Table 3 ($\Delta t = 8\times10^{-4}$). The second order accurate implicit multi-step scheme~\cite{Karniadakis1991} used in \cite{Stegmeir2023} likely has a smaller effective stability region than RK4. Next, we verify the correctness of the solution. A reference value $f=0.2734433$ is provided in \cite{Stegmeir2023} for the solution at $t=5$, $x=0.15\cos{2\pi/3}$, $y=0.15\sin{2\pi/3}$, $z=0$, computed using a 1D code. The absolute error found in our computation with respect to this value is reported in Table~\ref{tab:Screwpinch}, after removing $f_{\mathrm{bg},0}=5\times10^{-3}$. Since the grid is not aligned to the magnetic field, the values of $f(\rho=0.15,\theta=2\pi/3)$ are reconstructed with piecewise cubic Hermite interpolation of the simulation data. For the 4th order methods, we provide error values from $n_z=8$ and above, since the methods need at least 7 unique points per axis. Furthermore, notice that the interpolation error is expected to become comparable to the simulation error when $n_y\sim n_z$, since the Hermite interpolation and the parallel diffusion operator have 4th order accuracy.

\begin{table}
    \centering
    \begin{tabular}{ccccc}
   &  \multicolumn{2}{c}{Support Operator (SO)} & \multicolumn{2}{c}{Anti-Symmetry + SO (ASSO)}\\\hline
       $n_z$ & 2nd order & 4th order  & 2nd Order & 4th order \\\hline
         $4$ & \num{1.58e-1} &      --       & \num{1.55e-1} & -- \\ 
         $8$ & \num{1.09e-1} & \num{7.32e-2} & \num{7.75e-2} & \num{2.09e-2}\\
        $16$ & \num{4.66e-2} & \num{1.67e-2} & \num{3.01e-3} & \num{5.14e-3}\\
        $32$ & \num{7.70e-3} & \num{3.58e-4} & \num{1.30e-3} & \num{8.25e-5}\\
        $64$ & \num{8.02e-4} & \num{3.06e-4} & \num{9.83e-4} & \num{3.45e-4}\\
    \end{tabular}
    \caption{Estimated error $f-f_{\mathrm{ref}}$ from parallel diffusion simulations carried out in a screwpinch geometry are compared with the reference value from \cite{Stegmeir2023}. The value is reconstructed using piecewise cubic Hermite interpolation at $x=0.15\cos{2\pi/3}$, $y=0.15\sin{2\pi/3}$, $z=0$. }
    \label{tab:Screwpinch}
\end{table}

Snapshots of the final profiles are shown for the 2nd order methods, with $n_z=4,16,64$ are given in Fig.~\ref{fig:ScrewPinch}. The dashed red line indicates a  radius $\rho=0.15$, while the red dot indicates $(\rho,\theta)=(0.15,2\pi/3)$. The latter indicates the position where the diffusing lobe should cross the $x$--$y$ plane as it winds around the pinch. The SO2 simulations resemble previous published results \cite{Stegmeir2023}, including a gradual progression towards a higher quality as $n_z$ is increased. The ASSO2 simulations appear to be better resolved for a given $n_z$, with better defined and more concentric lobes as the field aligned structure winds and crosses $z=0$. It is also observed that the lobes approach their ideal return position more accurately. Indeed, looking at Table~\ref{tab:Screwpinch} it appears that ASSO2 achieves a similar level of accuracy as SO4 at the same resolution. A toroidal resolution of $n_z=16$ appears to be sufficient for both ASSO schemes to achieve an error below 1$\%$, while the SO methods need at least $n_z=32$. 

\section{Summary and conclusions}
\label{sec:Conclusion}
This work presents and tests two new discretization strategies for the parallel diffusion operator based on the anti-symmetry formulation. Starting from a simple diffusion equation, the physical observable -- a temperature, density, or pressure -- is expressed as a square quantity. We then employ the approach from Ref.~\cite{Halpern2018} to obtain an equation for its square root. The resulting expression recasts diffusion as a flow driven by the logarithmic gradient of the transported quantity, a non-linear operator. Since the operators involve successive applications of the first derivative operator, it is not possible to use a na\"ive, centered, finite difference scheme as in~\cite{Halpern2021}. Instead, some form of grid staggering is required. We present two discretization strategies that retain the anti-symmetry of the continuous space operator in discrete space. In the first strategy (``FV''), we evaluate vector components (fluxes) at their corresponding cell faces, reminiscent of finite volume methods. In the second strategy (``SO''), all 3 vector components are evaluated at the same corner of the cell. This method is inspired by the Support Operator method~\cite{Shashkov1996} employed to discretize the parallel Laplacian operator in \cite{Guenter2005,Guenter2007}.

A description of the new anti-symmetry formulation and its discretization are followed by qualitative analysis of the spectral properties of the discrete diffusion operators. The examination is itself qualitative because (a) the anti-symmetry operators are non-linear, and (b) the magnetic field must take a constant line pitch in order to keep the expressions tractable. Nevertheless, we derive 2D dispersion relations suggesting that the new anti-symmetry discretizations should possess an extended wave number range compared to numerical methods based on the traditional Laplace operator. This effect partially stems from the use of the square root of the observable, which maps profile gradients to a lower wave number. Thus, in principle, the new anti-symmetry numerical schemes should possess a higher effective resolution that traditional methods.

The same spectral analysis is carried out for the conventional parallel Laplace operator. Although it is known that standard finite differences result in considerable artificial diffusion, its causes are not, to our knowledge, described in existing literature. It is proposed that artificial diffusion originates from spectral inaccuracy of the discrete parallel Laplacian operator. The spectral inaccuracy manifests as a diffusion shortfall in isotropic diffusion problems, where it is generally tolerable. However, in the parallel diffusion problem, the overall spectral inaccuracy results in a loss of symmetry of the spectral response around the $k_\perp$--$k_\parallel$ axis. Hence, the effective diffusion at fixed $k_\parallel$ increases for large $k_\perp$. This effect is perceived as ``artificial diffusion'', which in fact results from small perpendicular scale structures diffusing faster than large perpendicular scale structures.

The wave number analysis carried out on the parallel Laplacian operators also clarifies why SO-based methods are perceived as better performing than FV-based methods. The improvement is attributed to the effects of a low-pass $k$-filter, which hides poor operator response at large $k_\perp$. However, the decrease in artificial perpendicular diffusion is achieved at the cost of worsening the parallel diffusion shortfall at large $k_\parallel$, \emph{i.e.} it hides the effect but does not address its underlying causes. This sometimes results in checkerboarding, in particular, in underresolved cases (Fig.~\ref{fig:Ring2Dx32}).

A central tenet of this work is that anti-symmetry-based methods possess a naturally extended effective resolution. Comprehensive verification exercises are carried out in 2D and 3D. The main conclusions of the numerical tests are as follows. All of the discretization methods (anti-symmetry and conventional) are found to achieve their asymptotic convergence order, which confirms their correct implementation. For a given accuracy order, the SO discretization performed significantly better than FV (roughly factor of $100$ improvement), while using the anti-symmetry strategy decreases the error by an additional factor of $2$ to $10$. Our hypothesis of an extended spectral range for the anti-symmetry methods appears to be confirmed.


Next, some additional observations are provided for each individual method. The conventional FV method performed, by far, the worst, with the infinity norm of the error (Fig.~\ref{fig:Convergence}) and the artificial perpendicular diffusion (Fig.~\ref{fig:EffectiveDiffusion}) remaining unacceptable even at high discretization order. The results shown here should discourage its use in any application where retaining the anisotropy of the transport tensor is important. The use of anti-symmetry (ASFV) reduces the amount of error considerably, yet not enough. The SO method performs well and produces small errors. However, in these numerical tests it sometimes led to solution quality issues: ringing, positivity violations, and (at low resolution) checkerboarding (\emph{e.g.}~Fig.~\ref{fig:Ring2Dx32}). These matters are ameliorated but do not disappear completely at higher order or higher resolution. This is a discouraging result for the SO method, since in practical settings numerical simulations are run at either marginal or insufficient resolution, and involve non-linear phenomena that generates cascading toward high wave numbers. It is suggested that codes using SO-based parallel diffusion method (a) be monitored for positivity violations and (b) be combined with a hyperdiffusion strategy to avoid checkerboarding.

The combined body of 2D and 3D simulation results presented suggests that the ASSO method be used wherever a large transport anisotropy must be reproduced. The method also shows good resolution up to 4 points per wavelength, \emph{e.g.} in the shallow pitch angle simulations, and is a good candidate for simulations including magnetic nulls. It produces negligible artificial perpendicular diffusion. The method works well even when the solution is underresolved, does not produce ringing, checkerboarding, or positivity violations.  Overall, we find that 2nd order ASSO is competitive with or better than 4th order SO, especially at lower resolutions. The major disadvantage of the ASSO method is that the entire system must be expressed and solved using the anti-symmetry strategy. Thus, it does not offer an effortless porting strategy for existing codes. Nonetheless, there might be additional advantages of using an anti-symmetry SO implementation of the plasma equations. We speculate that the improved spectral resolution might have a counterpart for the hyperbolic terms of the plasma equations. The plasma drift terms, in particular, revolve around the divergence of gradients oblique to the magnetic field. This issue will be explored in subsequent work describing the ASSO discretization of the perpendicular plasma dynamics. 

\section*{Acknowledgments}
The authors would like to thank B.~Zhu (LLNL) for providing the test case used in Section~\ref{sec:Simulations2D} along with its semi-analytical solution, Eq.~\ref{eq:Solution2D}. The analysis in that section partially follows B.~Zhu's  2024 Sherwood Fusion Theory Conference poster presentation. This research was partially supported by the U.S. Department of Energy, Office of Science, Office of Fusion Energy Sciences, Theory Program, under Award No. DE-FG02-95ER54309, and used resources of the National Energy Research Scientific Computing Center (NERSC), a Department of Energy Office of Science User Facility using NERSC award FES-ERCAP-27230.

This report was prepared as an account of work sponsored by an agency of the United States Government. Neither the United States Government nor any agency thereof, nor any of their employees, makes any warranty, express or implied, or assumes any legal liability or responsibility for the accuracy, completeness, or usefulness of any information, apparatus, product, or process disclosed, or represents that its use would not infringe privately owned rights. Reference herein to any specific commercial product, process, or service by trade name, trademark, manufacturer, or otherwise, does not necessarily constitute or imply its endorsement, recommendation, or favoring by the United States Government or any agency thereof. The views and opinions of authors expressed herein do not necessarily state or reflect those of the United States Government or any agency thereof.
\appendix

\section{Finite difference operators}\label{sec:Stencils}
A summary of the finite difference operators used in the manuscript is given below. The differential operators are denoted as $\dpd{}{}$ and the interpolation operators are denoted as $\dint{}{}$, with $^\pm$ superscripts denoting forward and backward grid transfers. The stencils are given for the $x$ direction with discrete index $i$ and grid spacing $\Delta x$. In a periodic domain, the backward grid transfer operator is related to the transpose of the forward transfer. This relationship is used to define the $^\pm$ operators together.
\subsection{Second order accurate stencils}
    \begin{align}
        \dint{x}{+} f & = \frac{f_{i+1} + f_i}{2} = \left(\dint{x}{-}\right)^T f, \\
        \dpd{x}{+} f & = \frac{f_{i+1}-f_i}{\Delta x}=\left(-\dpd{x}{-}\right)^T f.
    \end{align}
\subsection{Fourth order accurate stencils}
    \begin{align}
        \dint{x}{+} f & = \frac{-f_{i-1} + 9f_i + 9f_{i+1} -f_{i+2}}{16} = \left(\dint{x}{-}\right)^T f, \\
        \dpd{x}{+} f & = \frac{f_{i-1} - 27f_i + 27f_{i+1} -f_{i+2}}{24\Delta x} = -\left(\dpd{x}{-}\right)^T f.
    \end{align}
\subsection{Sixth order accurate stencils}
 \begin{align}
        \dint{x}{+} f & = \frac{3f_{i-2} -25f_{i-1} + 75f_i + 75f_{i+1} -25f_{i+2} + 3f_{i+2}}{256} 
        = \left(\dint{x}{-}\right)^T f, \\
        \dpd{x}{+} f & = \frac{1}{\Delta x}\left(
        -\frac{3}{640}f_{i-2} + \frac{25}{384}f_{i-1} -\frac{75}{64}f_{i} + \frac{75}{64}f_{i+1} - \frac{25}{384}f_{i+2} + \frac{3}{640}f_{i+3}
        \right) = -\left(\dpd{x}{-}\right)^T f.
    \end{align}

\bibliography{ALMA}

\end{document}

%% file: ListPackages.tex
\usepackage[margin = 1.2in]{geometry}
\usepackage[T1]{fontenc}
\usepackage[english]{babel}
\usepackage[latin1]{inputenc} 
\usepackage{mathtools}
\usepackage{siunitx}
\usepackage{amsmath}
\usepackage{amsfonts}
\usepackage{amsthm}
\usepackage{amssymb}
\usepackage{array}
\usepackage{algorithm} 
\usepackage{algorithmicx}
\usepackage{algpseudocode}
\usepackage{subcaption}
\usepackage{siunitx}
\usepackage{amsmath}
\usepackage{mathrsfs}
\usepackage{xpatch}

\usepackage{color}
\usepackage{url}
\usepackage{cleveref}
\usepackage{graphicx}
\usepackage{breakcites}
\usepackage{soul} 
\usepackage{enumitem}
\usepackage[title,titletoc,toc]{appendix}
\usepackage{rotating}

\newtheoremstyle{mytheoremstyle}{5pt}{5pt}{\itshape}{}{\bfseries}{.}{.5em}{} 
\theoremstyle{mytheoremstyle}

\newtheoremstyle{myremarkstyle}{3pt}{3pt}{\itshape}{}{\bfseries}{.}{.5em}{} 
\theoremstyle{myremarkstyle}

\usepackage{pifont}

%% file: headerCommands.tex
\let\baraccent=\= 
\renewcommand{\=}[1]{\stackrel{#1}{=}} 

\newcommand{\alma}{\texttt{ALMA}}

%% file: headerCommandsAlma.tex
\newcommand{\tensor}[1]{\overleftrightarrow{\mathbf{#1}}}
\newcommand{\mat}[1]{\mathbf{#1}}

\newcommand{\dgrad}{\mat{\Delta}}
\newcommand{\ddiv}{\mat{\Delta}^\dagger}
\newcommand{\dpd}[2]{\mat{D}^{#2}_{#1}}
\newcommand{\dint}[2]{\mat{I}^{#2}_{#1}}
\newcommand{\bhat}{\hat{\mathbf{b}}}
\newcommand{\xhat}{\hat{\mathbf{x}}}
\newcommand{\yhat}{\hat{\mathbf{y}}}

\newcommand{\divp}[1]{\nabla\cdot\left({#1}\right)}
\newcommand{\ddt}[2]{\frac{\partial#1}{\partial#2}}

\newcommand{\sqf}{\sqrt{f}}